# Tsallis non-extensive statistics, intermittent turbulence, SOC and chaos in the solar plasma. Part one: Sunspot dynamics

G.P. Pavlos[1], L.P. Karakatsanis[1], M.N. Xenakis[2]

*[1] Department of Electrical and Computer Engineering, Democritus University of Thrace, 67100 Xanthi, Greece*
*[2] German Research School for Simulation Sciences, Aachen, Germany*
*Email:* gpavlos@ee.duth.gr

## Abstract

*In this study, the nonlinear analysis of the sunspot index is embedded in the non-extensive statistical theory of Tsallis [5,7,8]. The $q-triplet$ of Tsallis, as well as the correlation dimension and the Lyapunov exponent spectrum were estimated for the SVD components of the sunspot index timeseries. Also the multifractal scaling exponent spectrum $f(a)$, the generalized Renyi dimension spectrum $D(q)$ and the spectrum $J(p)$ of the structure function exponents were estimated experimentally and theoretically by using the $q-entropy$ principle included in Tsallis non extensive statistical theory, following Arimitsu and Arimitsu [67,69]. Our analysis showed clearly the following: a) a phase transition process in the solar dynamics from high dimensional non Gaussian SOC state to a low dimensional non Gaussian chaotic state, b) strong intermittent solar turbulence and anomalous (multifractal) diffusion solar process, which is strengthened as the solar dynamics makes phase transition to low dimensional chaos in accordance to Ruzmakin, Zeleny and Milovanov studies [19,20,21] c) faithful agreement of Tsallis non equilibrium statistical theory with the experimental estimations of i) non-Gaussian probability distribution function $P(x)$, ii) multifractal scaling exponent spectrum $f(a)$ and generalized Renyi dimension spectrum $D_q$, iii) exponent spectrum $J(p)$ of the structure functions estimated for the sunspot index and its underlying non equilibrium solar dynamics.*

## 1. Introduction

The solar plasma dynamics is a prototype of non linearity and non integrability. According to Nilolis, Prigozine, Yonkov [1,2,3,4] it is known that far from equilibrium the nonlinear and non integrable dynamical systems can reveal novel characteristics as: non Gaussian self organized macroscopic dynamics with long range spatiotemporal correlations, anomalous and multifractal diffusion (intermittent turbulence), phase transitions and critical dynamics, as well as non Boltzmann – Gibbs statistical processes [5,6,7,8].

Thermodynamical theory is the fundamental theory of the macroscopic physics. As macroscopic phenomena must be related to the microscopic dynamics, the Boltzmann – Gibbs (BG) extensive statistical theory is a faithful theory for the state of thermodynamic equilibrium and the Gaussian dynamics. Moreover, the Tsallis non extensive statistical theory [6,7,8] is a novel faithful extension of the thermodynamics and its statistical interpretation for the far from equilibrium dynamics, where non Gaussian and long range correlation processes are developed. In this study, Tsallis non extensive statistical theory is used for the analysis/study of the solar plasma dynamics. Tsallis theory was already used by Kaniadakis et al [9] and Lavagno and Quarati [10] concerning the description of the solar core dynamics for the microscopic diffusion of electrons and ions, as well as the mechanics of the neutrino fluxes. For the outer regions of sun, sunspots and flare dynamics are two distinct but closely related topics of the solar activity. In the first part of this study we look at the sunspot dynamics by analyzing the sunspot index timeseries. The stochasticity of the sunspot process has been related theoretically with low dimensional dynamics, by Weiss [11] and Ruzmaikin [12]. Preliminary experimental evidence of solar chaos was given by Kurhts and Herzel [13] and Pavlos et al [14] by using nonlinear timeseries analysis. The solar low dimensional chaos has also been supported by many scientists [15,16,17,18]. The fractal properties of sunspots and their formation by fractal aggregates and clustering phenomena were studied by Zelenyi and Milovanov [19,20].

More than that, the anomalous diffusion and intermittent turbulence of the solar convection and photospheric motion was studied by Ruzmakin et al [21] and Cadavid et al [22] after using timeseries of solar Doppler images. In this direction Zimbatolo et al [23] studied numerically the magneto-hydrodynamic turbulence showing percolation, Levy flights and non-Gaussian dynamics. A modeling of the solar magnetic field at the photosphere and below, as a diffusion process in a non-Euclidean fractal environment, was introduced by Lawrence and Schrijver [24], and Petroway[25]. The magnetic turbulent character of solar plasma has been also related to self-organized critical process according to the theory of Bak [26] by many scientists [27,28,30,31]. The coexistence of SOC and low dimensional chaos as well as a phase transition process was supported by Karakatsanis and Pavlos [32] for the solar dynamics. The solar intermittent turbulence related with Tsallis non extensive statistics was also supported by Karakatsanis et al.[33].

  In this study, in relation with our previous results concerning the solar dynamics, we attempt a unifying description by using Tsallis non-extensive statistical theory in the non linear analysis of solar timeseries. In the first part of our study we present results concerning the analysis of sunspot index timeseries. In the second part (under preparation) we shall present similar results for the solar flare process. In the following sections, we summarize useful concepts of Tsallis theory (section 2). In section 3 we present the methodology of data analysis. In section 4 we show the results of sunspot index data analysis. Finally, in section 5 and 6 we summarize the highlights of our data analysis and discuss their physical meaning.

## 2. Theoretical Presuppositions for Data Analysis

In this section we summarize useful theoretical concepts underlying the methodology of data analysis. The Tsallis theory and the inner relation of this theory with solar turbulence are also presented. Next and according to these theoretical concepts, we describe the methodology of data analysis and the algorithm that was used for producing novel results, which are discussed in section 3.

### 2.1 Tsallis theory, turbulence and chaos

The statistical theory of Boltzmann and Gibbs (BG) is based on molecular chaos hypothesis known as "stosszahlansatz", which means the possibility of ergodic motion of the system in the microscopical phase space. That is the system can dynamically visit with equal probability all the allowed microscopic states. The dynamical attractor of the (BG) system dynamics can be a finite or an infinite dimensional chaotic object related with thermodynamical equilibrium. The probability distributions are Gaussians and the observed magnitudes (timeseries) reveal fluctuations in accordance with normal diffusion processes [8]. The equilibrium Gaussian random dynamics corresponds to physical states of uncorrelated or locally correlated noise. Timeseries signals related with this kind of dynamics, even if they are subjected to nonlinear distortion [34,35,36] revealing non-Gaussian and correlated profile, they can identified as Gaussian processes [37,38,39]. On the other side the far from equilibrium non linear dynamics can reveal strong self-organization and development of bifurcation processes causing long range correlations and strange attractors. Now the Gaussian statistics is unable to describe the fluctuations as these phenomena obey to non-Gaussian statistics, with break-down of the conditions of validity of the standard central limit theorem and the theorem of large numbers [1,5,8,40,41].

### 2.1.1 Tsallis entropy

The non-extensive generalization of the classical Gaussian Boltzmann-Gibbs statistical theory as it was done by Tsallis [5,7,8], showed the road for a deep unification of the microscopic and macroscopic physical theory, as well as for the physical interpretation of the non-Gaussian chaotic dynamics. Also, the Tsallis non-extensive statistical mechanics unifies the thermodynamical theory at equilibrium, with the non-equilibrium thermodynamics by extending the Boltzmann-Gibbs entropy to what is known, the last two-decades as Tsallis q-entropy desribe by the relations:

$$S_q = k \frac{1 - \sum_{i=1}^{N} p_i^q}{q-1}$$

$$= k \frac{1 - \int_{-\infty}^{+\infty} [p(x)] dx}{q-1}$$

(2.1)

where $p(i), (p(x))$ is the probability of occupation of state $(i)$ for discrete state space or $(x)$ for continuous state space and $q$ is a real number. This definition is based on $q$ - algebra:

$$e_q^x = [1 + (1-q)x]^{\frac{1}{1-q}},\tag{2.2}$$

where $e_q^x$ is the $q-$exponential function with inverse the $q-$logarithmic function:

$$\ln_q x = \frac{x^{1-q}-1}{1-q},\tag{2.3}$$

the $q-$logarithmic function satisfies the non-extensive relation:

$$\ln_q(x_A x_B) = \ln_q x_A + \ln_q x_B + (1-q)\ln_q x_A \ln_q x_B.\tag{2.4}$$

Tending to the limit $q \to 1$ the Tsallis entropy $S_q$ reproduces the Boltzmann-Gibbs entropy $S_{BG} = -k \sum p_i \ln p_i \lim_{x \to \infty}$. The optimal probability distribution $p(x)$ which optimizes the $S_q[p]$ function under appropriate constraints is given by the relation:

$$G_q(\beta;x) = \frac{\sqrt{\beta}}{C_q} e_q^{-\beta x^2},\tag{2.5}$$

while the normalizing constant $C_q$ is given as follows:

$$C_q = \begin{cases} \dfrac{2}{\sqrt{1-q}} \displaystyle\int_0^{\pi/2} (\cos t)^{\frac{3-q}{1-q}} dt = (2\sqrt{\pi}\,\Gamma(\frac{1}{1-q}))/((3-q)\sqrt{1-q}\,\Gamma(\frac{3-q}{2(1-q)})), & -\infty < q < 1, \\[3mm] \sqrt{\pi}, & q = 1, \\[3mm] \dfrac{2}{\sqrt{q-1}} \displaystyle\int_0^{\infty} (1+y^2)^{\frac{-1}{q-1}} dy = (\sqrt{\pi}\,\Gamma(\frac{3-q}{2(q-1)}))/(\sqrt{q-1}\,\Gamma(\frac{1}{q-1})), & 1 < q < 3, \end{cases}\tag{2.6}$$

The function (2.5) maximize (minimize) for $q > 0$ $(q < 0)$ the functional $S_q[p]$ under appropriate constraints [8,41]. The above solutions known as $q-$Gaussians are solutions also of equations corresponding to anomalous diffusion processes including: Levy anomalous diffusion, correlated anomalous diffusion and generalized fractal Fokker-Planck equations [8,42,43,44].

### 2.1.2 Anomalous Diffusion

Generally the fluctuations of observed magnitudes in the form of timeseries $\{x(t)\}$ can be explained as caused by diffusion processes in the physical or the phase space of the system. Chaotic motion in the physical or the phase space (state) causes exponential or other kinds of growing distances between initially close orbits. When the orbit distance grows exponentially with time as:

$$w(t) = w(0)\exp^{\lambda t},\tag{2.7}$$

where $\lambda$ is a positive Lyapunov exponent, we observe a rapid mixing of orbits within the time interval $\tau = 1/\lambda$. Then kinetic equations arise as coarse graining and averaging at timescales grater than the mixing time $\tau = 1/\lambda$. The well-known Gaussian and Poissonian chaotic dynamics with space and temporal diffusion can be produced from chaotic dynamics [45,46,47,48,49]. Anomalous diffusion processes include long range correlations with mean-squared jump distances:

$$\left\langle x^2(t) \right\rangle \sim t^\gamma,$$
(2.8)

with $\gamma \neq 1$, ($\gamma = 1$ corresponds to normal Gaussian diffusion processes) while $\gamma > 1$ ($\gamma < 1$) corresponds to super diffusion (sub diffusion). Anomalous diffusion processes are caused by chaotic motions on multifractal and multiscale self-similar set of states in physical or state space [48,50,51]. Moreover, anomalous diffusion can be caused by strong nonlinear and chaotic dynamics, which creates multifractal attracting structures in state spaces or multifractal materials structures in which the diffusion process happens by the fractality of space and time by themselves, [52,53,54].

Generally the anomalous diffusion processes described by fractional Fokker-Planck equations include Levy distributions, which asymptotically are related to $q$ – Gaussians of Tsallis statistics as follows:

$$L_\gamma(x) \sim \frac{1}{|x|^{1+\gamma}}, \quad |x| \to \infty \quad 0 < \gamma < 2$$
(2.9a)

$$P_q(x) \sim \frac{1}{|x|^{\frac{2}{q-1}}}, \quad |x| \to \infty \quad 1 < q < 3.$$
(2.9b)

where $L_\gamma(x), P_q(x)$ correspond to Levy and $q$ – Gaussian distributions.

These relations indicate that:

$$\gamma = \begin{cases} 2, & if \quad q \leq \frac{5}{3} \\ \dfrac{3-q}{q-1}, & if \quad \frac{5}{3} < q < 3 \end{cases}$$
(2.10)

### 2.1.3 Long Range Correlations

According to [7] the $q$ – statistics for $q \neq 1$ indicates the existence of long range correlations as we can conclude by the relation:

$$S_q(A+B) = S_q(A) + S_q(B/A) + (1-q)S_q(A)S_q(B/A) = S_q(B) + S_q(A/B) + (1-q)S_q(B)S_q(A/B),$$
(2.11)

where $A, B$ are subsystems of the system $A+B$ and $S_q(A/B), S_q(B/A)$ correspond to intercorrelations of subsystems $A$ and $B$ when they are statistically dependent and statistically correlated according to the relation:

$$P_{ij}^{A+B} \neq P_i^A P_j^B$$
(2.12)

When the probabilities $P_{ij}^{A+B} = P_i^A P_j^B$, indicate that the subsystems are statistically independent and uncorrelated and the relation (2.11) is transformed to :

$$S_q(A+B) = S_q(A) + S_q(B) + (1-q)S_q(A)S_q(B).$$
(2.13)

From (2.13) it becomes clear the non-extensive character of $q$ – entropy, even at the case of Gaussian ($q = 1$) and internally uncorrelated processes.

### 2.1.4 Chaotic attractors and Self-Organization

When the dynamics is strongly nonlinear then for the far from equilibrium processes it is possible to be created strong self-organization and intensive reduction of dimensionality of the state space, by an attracting low dimensional set with parallel development of long range correlations in space and time. The attractor can be periodic (limit cycle, limit m-torus), simply chaotic (mono-fractal) or strongly chaotic with multiscale and multifractal profile as well as attractors with weak chaotic profile known as SOC states. This spectrum of distinct dynamical profiles can be obtained as distinct critical points (critical states) of the nonlinear dynamics, after successive bifurcations as the control parameters change. The fixed points can be estimated by using a far from equilibrium renormalization process as it was indicated by Chang [55,56].

From this point of view phase transition processes can be developed by between different critical states, when the order parameters of the system are changing. The far from equilibrium development of chaotic (weak or strong) critical states include long range correlations and multiscale internal self organization. Now, these far from equilibrium self organized states, the equilibrium BG statistics and BG entropy, are transformed and replaced by the Tsallis extension of $q$ – statistics and Tsallis entropy. The extension of renormalization group theory and critical dynamics, under the $q$ – extension of partition function, free energy and path integral approach has been also indicated [57,58,59,60]. The multifractal structure of the chaotic attractors can be described by the generalized Rényi fractal dimensions:

$$D_{\bar{q}} = \frac{1}{\bar{q}-1} \lim_{\lambda \to 0} \frac{\log \sum_{i=1}^{N\lambda} p_i^{\bar{q}}}{\log \lambda}, \qquad (2.14)$$

where $p_i \sim \lambda^{\alpha(i)}$ is the local probability at the location ($i$) of the phase space, $\lambda$ is the local size of phase space and $a(i)$ is the local fractal dimension of the dynamics. The Rényi $\bar{q}$ numbers (different from the $q$ – index of Tsallis statistics) take values in the entire region ($-\infty, +\infty$) of real numbers. The spectrum of distinct local fractal dimensions $\alpha(i)$ is given by the estimation of the function $f(\alpha)$ [61,62] for which the following relations hold:

$$\sum p_i^{\bar{q}} = \int d\alpha' p(\alpha') \lambda^{-f(\alpha')} d\alpha' \qquad (2.15)$$

$$\tau(\bar{q}) \equiv (\bar{q}-1)D\bar{q} \overset{a}{=} \bar{q}\alpha - f(\alpha) \qquad (2.16)$$

$$a(\bar{q}) = \frac{d[\tau(\bar{q})]}{d\bar{q}} \qquad (2.17)$$

$$f(\alpha) = \bar{q}\alpha - \tau(\bar{q}), \qquad (2.18)$$

The physical meaning of these magnitudes included in relations (2.15-2.18) can be obtained if we identify the multifractal attractor as a thermodynamical object, where its temperature ($T$), free energy ($F$), entropy ($S$) and internal energy ($U$) are related to the properties of the multifractal attractor as follows:

$$\left.\begin{array}{ll} \bar{q} \Rightarrow \dfrac{1}{T}, & \tau(\bar{q}) = (\bar{q}-1)D_q \Rightarrow F \\[2mm] \alpha \Rightarrow U, & f(\alpha) \Rightarrow S \end{array}\right\} \qquad (2.19)$$

This correspondence presents the relations (2.17 -2.19) as a thermodynamical Legendre transform [63]. When $\bar{q}$ increases to infinite ($+\infty$), which means, that we freeze the system ($T_{(q=+\infty)} \to 0$), then the trajectories (fluid lines) are closing on the attractor set, causing large probability values at regions of low fractal dimension, where $\alpha = \alpha_{\min}$ and $D_{\bar{q}} = D_{-\infty}$. Oppositely, when $\bar{q}$ decreases to infinite ($-\infty$), that is we warm up the system ($T_{(q=-\infty)} \to 0$) then the trajectories are spread out at regions of high fractal dimension ($\alpha \Rightarrow \alpha_{\max}$). Also for $\bar{q}' > \bar{q}$ we have $D_{\bar{q}'} < D_{\bar{q}}$ and $D_{\bar{q}} \Rightarrow D_{+\infty}(D_{-\infty})$ for $\alpha \Rightarrow \alpha_{\min}(\alpha_{\max})$ correspondingly. However, the above description presents only a weak or limited analogy between multifractal and thermodynamical objects. The real thermodynamical character of the multifractal objects and multiscale dynamics was discovered after the definition by Tsallis [5] of the $q - $entropy related with the $q - $statistics as it was summarized previously in relations (2.1-2.13).

## 2.1.5 Intermittent Turbulence

According to previous description dissipative nonlinear dynamics can produce self-organization and long range correlations in space and time. In this case we can imagine the mirroring relationship between the phase space multifractal attractor and the corresponding multifractal turbulence dissipation process of the dynamical system in the physical space. Multifractality and multiscaling interaction, chaoticity and mixing or diffusion (normal or anomalous), all of them can be manifested in both the state (phase) space and the physical (natural) space as the mirroring of the same complex dynamics. We could say that turbulence is for complexity theory, what the blackbody radiation was for quantum theory, as all previous characteristics can be observed in turbulent states. The theoretical description of turbulence in the physical space is based upon the concept of the invariance of the HD or MHD equations upon scaling transformations to the space-time variables ($\vec{X}, t$) and velocity ($\vec{U}$):

$$\vec{X}' = \lambda \vec{X}, \ \vec{U}' = \lambda^{\alpha/3} \vec{U}, t' = \lambda^{1-\alpha/3} t \qquad (2.20)$$

and corresponding similar scaling relations for other physical variables [64,65,66]. Under these scale transformations the dissipation rate of turbulent kinetic or dynamical field energy $E_n$ (averaged over a scale $l_n = l_o \delta_n = R_o \delta_n$) rescales as $\varepsilon_n$:

$$\varepsilon_n \sim \varepsilon_0 (l_n \backslash l_0)^{\alpha - 1} \qquad (2.21)$$

Kolmogorov [41] assumes no intermittency as the locally averaged dissipation rate, in reality a random variable, is independent of the averaging domain. This means in the new terminology of Tsallis theory that Tsallis $q$-indices satisfy the relation $q = 1$ for the turbulent dynamics in the three dimensional space. That is the multifractal (intermittency) character of the HD or the MHD dynamics consists in supposing that the scaling exponent $\alpha$ included in relations (2.20, 2.21) takes on different values at different interwoven fractal subsets of the $d - $dimensional physical space in which the dissipation field is embedded. The exponent $\alpha$ and for values $a < d$ is related with the degree of singularity in the field's gradient ($\frac{\partial A(x)}{\partial x}$) in the $d - $dimensional natural space [67]. The gradient singularities cause the anomalous diffusion in physical or in phase space of the dynamics. The total dissipation occurring in a $d - $dimensional space of size $l_n$ scales also with a global dimension $D_{\bar{q}}$ for powers of different order $\bar{q}$ as follows:

$$\sum_n \varepsilon_n^{\bar{q}} l_n^{\ d} \sim l_n^{\ (\bar{q}-1)D_{\bar{q}}} = l_n^{\ \tau(\bar{q})} \qquad (2.22)$$

Supposing that the local fractal dimension of the set $dn(a)$ which corresponds to the density of the scaling exponents in the region ($\alpha, \alpha + d\alpha$) is a function $f_d(a)$ according to the relation:

$$dn(\alpha) \sim \ln^{-f_d(\alpha)} da \qquad (2.23)$$

where $d$ indicates the dimension of the embedding space, then we can conclude the Legendre transformation between the mass exponent $\tau(\overline{q})$ and the multifractal spectrum $f_d(a)$:

$$\left.\begin{array}{l} f_d(a) = a\overline{q} - (\overline{q}-1)(D_{\overline{q}} - d + 1) + d - 1 \\[2mm] a = \dfrac{d}{d\overline{q}}[(\overline{q}-1)(D_{\overline{q}} - d + 1)] \end{array}\right\} \qquad (2.24)$$

For linear intersections of the dissipation field, that is $d = 1$ the Legendre transformation is given as follows:

$$f(a) = a\overline{q} - \tau(\overline{q}), \;\; a = \frac{d}{d\overline{q}}[(q-1)D_q] = \frac{d}{d\overline{q}}\tau(\overline{q}), \overline{q} = \frac{df(a)}{da} \qquad (2.25)$$

The relations (2.23-2.25) describe the multifractal and multiscale turbulent process in the physical state. The relations (2.14-2.19) describe the multifractal and multiscale process on the attracting set of the phase space. From this physical point of view, we suppose the physical identification of the magnitudes $D_{\overline{q}}, a, f(a)$ and $\tau(\overline{q})$ estimates in the physical and the corresponding phase space of the dynamics. By using experimental timeseries we can construct the function $D_{\overline{q}}$ of the generalized Rényi $d-$ dimensional space dimensions, while the relations (2.24) allow the calculation of the fractal exponent ($a$) and the corresponding multifractal spectrum $f_d(a)$. For homogeneous fractals of the turbulent dynamics the generalized dimension spectrum $D_{\overline{q}}$ is constant and equal to the fractal dimension, of the support [64]. Kolmogorov [68] supposed that $D_{\overline{q}}$ does not depend on $\overline{q}$ as the dimension of the fractal support is $D_q = 3$. In this case the multifractal spectrum consists of the single point ($a = 1$ and $f(1) = 3$). The singularities of degree ($a$) of the dissipated fields, fill the physical space of dimension $d$ with a fractal dimension $F(a)$, while the probability $P(a)da$, to find a point of singularity ($a$) is specified by the probability density $P(a)da \sim \ln^{d - F(a)}$. The filling space fractal dimension $F(a)$ is related with the multifractal spectrum function $f_d(a) = F(a) - (d-1)$, while according to the distribution function $\Pi_{dis}(\varepsilon_n)$ of the energy transfer rate associated with the singularity $a$ it corresponds to the singularity probability as $\Pi_{dis}(\varepsilon_n)d\varepsilon_n = P(a)da$ [67].

Moreover the partition function $\sum_i P_i^{\overline{q}}$ of the Rényi fractal dimensions estimated by the experimental timeseries includes information for the local and global dissipation process of the turbulent dynamics as well as for the local and global dynamics of the attractor set, as it is transformed to the partition function $\sum_i P_i^q = Z_q$ of the Tsallis $q-$statistical theory.

In the following, we follow Arimitsu and Arimitsu [67,69] for the theoretical estimation of significant quantitative relations which can also be estimated experimentally. The probability singularity distribution $P(a)$ can be estimated as extremizing the Tsallis entropy functional $S_q$. According to Arimitsu and Arimitsu [67] the extremizing probability density function $P(a)$ is given as a $q-$exponential function:

$$P(a) = Z_q^{-1}[1-(1-q)\frac{(a-a_0)^2}{2X/\ln 2}]^{\frac{1}{1-q}} \tag{2.26}$$

where the partition function $Z_q$ is given by the relation:

$$Z_q = \sqrt{2X/[(1-q)\ln 2]}\ B(1/2, 2/1-q), \tag{2.27}$$

and $B(a,b)$ is the Beta function. The partition function $Z_q$ as well as the quantities $X$ and $q$ can be estimated by using the following equations:

$$\left.\begin{array}{l} \sqrt{2X} = \left[\sqrt{a_0^2+(1-q)^2}-(1-q)\right]/\sqrt{b} \\ b = (1-2^{-(1-q)})/[(1-q)\ln_2] \end{array}\right\} \tag{2.28}$$

We can conclude for the exponent's spectrum $f(a)$ by using the relation $P(a) \approx \ln^{d-F(a)}$ as follows:

$$f(a) = D_0 + \log_2[1-(1-q)\frac{(a-a_o)^2}{2X/\ln 2}]/(1-q)^{-1} \tag{2.29}$$

where $a_0$ corresponds to the $q-$expectation (mean) value of $a$ through the relation:

$$<(a-a_0)^2>_q = (\int da P(a)^q (a-a_0)^q)/\int da P(a)^q. \tag{2.30}$$

while the $q-$expectation value $a_0$ corresponds to the maximum of the function $f(a)$ as $df(a)/da \,|\, a_0 = 0$. For the Gaussian dynamics $(q \to 1)$ we have mono-fractal spectrum $f(a_0) = D_0$. The mass exponent $\tau(\overline{q})$ can be also estimated by using the inverse Legendre transformation: $\tau(\overline{q}) = a\overline{q} - f(a)$ (relations 2.24 − 2.25) and the relation (2.29) as follows:

$$\tau(\overline{q}) = \overline{q}a_0 - 1 - \frac{2X\overline{q}^2}{1+\sqrt{C_{\overline{q}}}} - \frac{1}{1-q}[1-\log_2(1+\sqrt{C_{\overline{q}}})], \tag{2.31}$$

Where $C_{\overline{q}} = 1 + 2\overline{q}^2(1-q)X\ln 2$.

The relation between $a$ and $q$ can be found by solving the Legendre transformation equation $\overline{q} = df(a)/da$. Also if we use the equation (2.29) we can obtain the relation:

$$a_{\overline{q}} - a_0 = (1-\sqrt{C_{\overline{q}}})/[\overline{q}(1-q)\ln 2] \tag{2.32}$$

The $q-$index is related to the scaling transformations (2.20) of the multifractal nature of turbulence according to the relation $q = 1 - a$. Arimitsu and Arimitsu [69] estimated the $q-$index by analyzing the fully developed turbulence state in terms of Tsallis statistics as follows:

$$\frac{1}{1-q} = \frac{1}{a_-} - \frac{1}{a_+} \tag{2.33}$$

where $a_{\pm}$ satisfy the equation $f(a_{\pm}) = 0$ of the multifractal exponents spectrum $f(a)$. This relation can be used for the estimation of $q_{sen}-$index included in the Tsallis $q-$triplet (see next section).

The above analysis based at the extremization of Tsallis entropy can be also used for the theoretical estimation of the structure functions scaling exponent spectrum $J(p)$ of the $S_p(\tau)$, where $p = 1, 2, 3, 4, \ldots$ The structure functions were first introduced by Kolmogorov [68] defined as statistical moments of the field increments:

$$S_p(\vec{r}) = <\mid u(\vec{x} + \vec{d}) - u(\vec{x}) \mid^p > = <\mid \delta u_n \mid^p > \tag{2.34}$$

$$S_p(\vec{r}) = <\mid u(\vec{x} + \Delta\vec{x}) - u(\vec{x}) \mid^p > \tag{2.35}$$

After discretization of $\Delta\vec{x}$ displacement the above relation can be identified to:

$$Sp(l^n) = <\mid \delta u_n \mid^p > \tag{2.36}$$

The field values $u(\vec{x})$ can be related with the energy dissipation values $\varepsilon_n$ by the general relation $\varepsilon_n = (\delta u_n)^3 / l^n$ in order to obtain the structure functions as follows:

$$S_p(l^n) = <(\varepsilon_n / \varepsilon_0)^p > = <\delta_n^{p(a-1)} > = \delta_n^{j(p)} \tag{2.37}$$

where the averaging processes $< \ldots >$ is defined by using the probability function $P(a)da$ as $< \ldots > = \int da (\ldots) P(a)$. By this, the scaling exponent $J(p)$ of the structure functions is given by the relation:

$$J(p) = 1 + \tau(\overline{q} = \frac{p}{3}) \tag{2.38}$$

By following Arimitsu [67,69] the relation (2.30) leads to the theoretical prediction of $J(p)$ after extremization of Tsallis entropy as follows:

$$J(p) = \frac{a_0 p}{3} - \frac{2Xp^2}{q(1 + \sqrt{C_{p/3}})} - \frac{1}{1-q}[1 - \log_2(1 + \sqrt{C_{p/3}})] \tag{2.39}$$

The first term $a_0 p / 3$ corresponds to the original of known Kolmogorov theory (K41) according to which the dissipation of field energy $\varepsilon_n$ is identified with the mean value $\varepsilon_0$ according to the Gaussian self-similar homogeneous turbulence dissipation concept, while $a_0 = 1$ according to the previous analysis for homogeneous turbulence. According to this concept the multifractal spectrum consists of a single point. The next terms after the first in the relation (2.39) correspond to the multifractal structure of intermittence turbulence indicating that the turbulent state is not homogeneous across spatial scales. That is, there is a greater spatial concentration of turbulent activity at smaller than at larger scales. According to Abramenko [70] the intermittent multifractal (inhomogeneous) turbulence is indicated by the general scaling exponent $J(p)$ of the structure functions according to the relation:

$$J(p) = \frac{p}{3} + T^{(u)}(p) + T^{(F)}(p),$$ (2.40)

where the $T^{(u)}(p)$ term is related with the dissipation of kinetic energy and the $T^{(F)}(p)$ term is related to other forms of field's energy dissipation as the magnetic energy at MHD turbulence [70,71].

The scaling exponent spectrum $J(p)$ can be also used for the estimation of the intermittency exponent $\mu$ according to the relation:

$$S(2) \equiv <\varepsilon^2 / \varepsilon > \sim \delta_n^{\mu} = \delta_n^{J(2)}$$ (2.41)

from which we conclude that $\mu = J(2)$. The intermittency turbulence correction to the law $P(f) \sim f^{-5/3}$ of the energy spectrum of Kolmogorov's theory is given by using the intermittency exponent:

$$P(f) \sim f^{-(5/3+\mu)}$$ (2.42)

The previous theoretical description can be used for the theoretical interpretation of the experimentally estimated structure function, as well as for relating physically the results of data analysis with Tsallis statistical theory, as it is described in the next sections.

### 2.1.6 The $q-$Triplet of Tsallis theory

The non-extensive statistical theory is based mathematically on the nonlinear equation:

$$\frac{dy}{dx} = y^q, \; ( \; y(0) = 1, q \in \Re \; )$$ (2.43)

with solution the $q-$exponential function defined previously in equation (2.2). The solution of this equation can be realized in three distinct ways included in the $q-$triplet of Tsallis: ($q_{sen}, q_{stat}, q_{rel}$). These quantities characterize three physical processes which are summarized here, while the $q-$triplet values characterize the attractor set of the dynamics in the phase space of the dynamics and they can change when the dynamics of the system is attracted to another attractor set of the phase space. The equation (2.36) for $q = 1$ corresponds to the case of equilibrium Gaussian Boltzmann-Gibbs (BG) world [7,8]. In this case of equilibrium BG world the $q-$triplet of Tsallis is simplified to ($q_{sen} = 1, q_{stat} = 1, q_{rel} = 1$).

### a. The $q_{stat}$ index and the non-extensive physical states.

According to [7,8] the long range correlated metaequilibrium non-extensive physical process can be described by the nonlinear differential equation:

$$\frac{d(p_i Z_{stat})}{dE_i} = -\beta q_{stat} (p_i Z_{stat})^{q_{stat}}$$ (2.44)

The solution of this equation corresponds to the probability distribution:

$$p_i = e_{q_{stat}}^{-\beta_{stat} E_i} / Z_{q_{stat}}$$ (2.45)

where $\beta_{q_{stat}} = \dfrac{1}{KT_{stat}}$, $Z_{stat} = \sum_j e_{q_{stat}}^{-\beta q_{stat} E_j}$.

Then the probability distribution function is given by the relations:

$$p_i \propto \left[1-(1-q)\beta_{q_{stat}} E_i\right]^{1/1-q_{stat}} \qquad (2.46)$$

for discrete energy states $\{E_i\}$ by the relation:

$$p(x) \propto \left[1-(1-q)\beta_{q_{stat}} x^2\right]^{1/1-q_{stat}} \qquad (2.47)$$

for continuous $X$ states $\{X\}$, where the values of the magnitude $X$ correspond to the state points of the phase space.

The above distributions functions (2.46, 2.47) correspond to the attracting stationary solution of the extended (anomalous) diffusion equation related with the nonlinear dynamics of system [8]. The stationary solutions $P(x)$ describe the probabilistic character of the dynamics on the attractor set of the phase space. The non-equilibrium dynamics can be evolved on distinct attractor sets depending upon the control parameters values, while the $q_{stat}$ exponent can change as the attractor set of the dynamics changes.

**b. The $q_{sen}$ index and the entropy production process**

The entropy production process is related to the general profile of the attractor set of the dynamics. The profile of the attractor can be described by its multifractality as well as by its sensitivity to initial conditions. The sensitivity to initial conditions can be described as follows:

$$\frac{d\xi}{d\tau} = \lambda_1 \xi + (\lambda_q - \lambda_1)\xi^q \qquad (2.48)$$

where $\xi$ describes the deviation of trajectories in the phase space by the relation: $\xi \equiv \lim_{\Delta(x)\to 0}\{\Delta x(t) \setminus \Delta x(0)\}$ and $\Delta x(t)$ is the distance of neighboring trajectories [72]. The solution of equation (2.41) is given by:

$$\xi = \left[1 - \frac{\lambda q_{sen}}{\lambda_1} + \frac{\lambda q_{sen}}{\lambda_1} e^{(1-q_{sen})\lambda_1 t}\right]^{\frac{1}{1-q}} \qquad (2.49)$$

The $q_{sen}$ exponent can be also related with the multifractal profile of the attractor set by the relation:

$$\frac{1}{q_{sen}} = \frac{1}{a_{min}} - \frac{1}{a_{max}} \qquad (2.50)$$

where $a_{min}(a_{max})$ corresponds to the zero points of the multifractal exponent spectrum $f(a)$ [8,67,72]. That is $f(a_{min}) = f(a_{max}) = 0$.

The deviations of neighboring trajectories as well as the multifractal character of the dynamical attractor set in the system phase space are related to the chaotic phenomenon of entropy production according to Kolmogorov – Sinai entropy production theory and the Pesin theorem [8]. The $q$ – entropy production is summarized in the equation:

$$K_q \equiv \lim_{t\to\infty}\lim_{W\to\infty}\lim_{N\to\infty}\frac{<S_q>(t)}{t}. \qquad (2.51)$$

The entropy production ($dS_q / t$) is identified with $K_q$, as $W$ are the number of non-overlapping little windows in phase space and $N$ the state points in the windows according to the relation $\sum_{i=1}^{W} N_i = N$. The $S_q$ entropy is estimated by the probabilities $P_i(t) \equiv N_i(t) / N$. According to Tsallis the entropy production $K_q$ is finite only for $q = q_{sen}$ [8,72].

### c. The $q_{rel}$ index and the relaxation process

The thermodynamical fluctuation − dissipation theory [73] is based on the Einstein original diffusion theory (Brownian motion theory). Diffusion process is the physical mechanism for extremization of entropy. If $\Delta S$ denote the deviation of entropy from its equilibrium value $S_0$, then the probability of the proposed fluctuation that may occur is given by:

$$P \sim \exp(\Delta s / k) \, . \qquad (2.52)$$

The Einstein − Smoluchowski theory of Brownian motion was extended to the general Fokker − Planck diffusion theory of non-equilibrium processes. The potential of Fokker − Planck equation may include many metaequilibrium stationary states near or far away from the basic thermodynamical equilibrium state. Macroscopically, the relaxation to the equilibrium stationary state can be described by the form of general equation as follows:

$$\frac{d\Omega}{d\tau} \simeq -\frac{1}{\tau}\Omega \, , \qquad (2.53)$$

where $\Omega(t) \equiv [O(t) - O(\infty)] / [O(0) - O(\infty)]$ describes the relaxation of the macroscopic observable $O(t)$ relaxing towards its stationary state value. The non-extensive generalization of fluctuation − dissipation theory is related to the general correlated anomalous diffusion processes [8]. Now, the equilibrium relaxation process (2.46) is transformed to the metaequilibrium non-extensive relaxation process:

$$\frac{d\Omega}{dt} = -\frac{1}{T_{q_{rel}}}\Omega^{q_{rel}} \qquad (2.54)$$

the solution of this equation is given by:

$$\Omega(t) \simeq e_{q_{rel}}^{-t/\tau_{rel}} \qquad (2.55)$$

The autocorrelation function $C(t)$ or the mutual information $I(t)$ can be used as candidate observables $\Omega(t)$ for the estimation of $q_{rel}$. However, in contrast to the linear profile of the correlation function, the mutual information includes the non linearity of the underlying dynamics and it is proposed as a more faithful index of the relaxation process and the estimation of the Tsallis exponent $q_{rel}$.

### 2.2 Methodology of Data Analysis

In this study we use the method of singular value decomposition (SVD) in order to uncover the hidden solar dynamics, underlying the sunspot index timeseries. Moreover the non extensivity of the solar dynamics is studied in relation to the turbulence and chaotic profiles of the solar plasma convection.

### 2.2.1 Singular Value Decomposition (SVD) Analysis

The embedding theory of Takens [74] and the theory of SVD analysis can be used for the discrimination of deterministic and noisy (stochastic) components included in the observed signals, as well as for the discrimination of distinct dynamical components [39,75,76,77].

The trajectory matrix ($X$) obtained by the reconstructed trajectory in the embedding state space can be analysed in three matrices ($S, \Sigma, C$) according to the relation:

$$X = S\Sigma C^T \qquad (2.56)$$

The singular vectors $\{\vec{c}_i\}$ and singular values $\{\sigma_i\}$ of the covariance matrix $X^T X$ can be used for the extraction of the deterministic components of the dynamics underlying to the experimental signal as follows:

$$X(t) = \sum_{i=1}^{m} V_i(t) . \qquad (2.57)$$

The $V_i(t)$ time series (known as SVD components) are given for the first column of the matrix $(XC_i)C_i^T$, corresponding to the first non-zero eigenvalues or the eigenvalues above the noise level according to the relation:

$$\sigma_1 \geq \sigma_2 \geq ... \geq \sigma_n > \sigma_{n+k} , \qquad (2.58)$$

where $\sigma_{n+k} = \begin{cases} 0 \\ cons \tan t \end{cases}, k = 1, 2, ... m - n$.

Distinct dynamical components can be discriminated between the first ($V_1$) and higher SVD components ($V_{2-n}$), as the singular values correspond to the lengths of the principal evolves in the subspace spanned by the eigenvectors $\{c_i\}$ in the reconstructed state phase.

## 2.2.2 Flatness Coefficient $F$

The intermittent nature of the magnetospheric plasma dynamics can be investigated through the Probability Density Functions (PDF) of a set of two-point differenced time series of an original time series $\delta B_\tau(t) = B(t + \tau) - B(t)$ which can be any physical quantity. The coefficient $F$ corresponding to the flatness values of the two-point difference for the observed time series is defined as:

$$F = \frac{< \delta B_\tau(t)^4 >}{< \delta B_\tau(t)^2 >^2} \qquad (2.59)$$

The coefficient $F$ for a Gaussian process is equal to 3. Deviation from Gaussian distributions implies intermittency, while the parameter $\tau$ represents the spatial size of the plasma eddies, which contribute to the energy cascade process.

According to general theory of turbulence, intermittency appears in the heavy tails of the distribution functions as the dynamics in the vortex is non-random, but deterministic.

## 2.2.3 The $q$ – triplet estimation

According to previous analysis concerning the $q$ – triplet of Tsallis, we estimate the ($q_{sen}, q_{stat}, q_{rel}$) as follows:

The $q_{sen}$ index is given by the relation:

$$q_{sen} = 1 + \frac{a_{max} a_{min}}{a_{max} - a_{min}} .$$ (2.60)

The $a_{max}, a_{min}$ values correspond to the zeros of multifractal spectrum function $f(a)$, which is estimated by the Legendre transformation $f(a) = \bar{q}a - (\bar{q} - 1)D_{\bar{q}}$, where $D_{\bar{q}}$ describes the Rényi generalized dimension of the sunspot signals in accordance to relation:

$$D_{\bar{q}} = \frac{1}{q-1} \cdot \lim \left( \frac{\log \sum p_i^q}{\log r} \right) \text{ for } r \to 0 .$$ (2.61)

The $f(a)$ and $D(\bar{q})$ functions can be estimated experimentally by using the relations (2.14-2.18) and the underlying theory according to March and Tu [71] and Consolini [78]. The same functions can be also estimated theoretically by using the Tsallis $q-$entropy principle according to the relations (2.24, 2.25) and (2.29, 2.31) of section (2.1.5).

The q$_{stat}$ values are derived from the observed Probability Distribution Functions (PDF) according to the Tsallis q-exponential distribution:

$$PDF[\Delta Z] \equiv A_q \left[ 1 + (q-1)\beta_q (\Delta Z)^2 \right]^{\frac{1}{1-q}},$$ (2.62)

where the coefficient A$_q$, β$_q$ denote the normalization constants and $q \equiv q_{stat}$ is the entropic or non-extensivity factor ($q_{stat} \leq 3$) related to the size of the tail in the distributions. Our statistical analysis is based on the algorithm as described in [79]. We construct the $PDF[\Delta Z]$ which is associated to the first difference $\Delta z = z_{n+1} - z_n$ of the experimental sunspot time series, while the $\Delta Z$ range is subdivided into little ``cells'' (data binning process) of width $\delta z$, centered at $z_i$ so that one can assess the frequency of $\Delta z$ -values that fall within each cell/bin. The selection of the cell-size $\delta z$ is a crucial step of the algorithmic process and its equivalent to solving the binning problem: a proper initialization of the bins/cells can speed up the statistical analysis of the data set and lead to a convergence of the algorithmic process towards the exact solution. The resultant histogram is being properly normalized and the estimated q-value corresponds to the best linear fitting to the graph $\ln_q(p(z_i)) \, vs \; z_i^2$, where $\ln_q$ is the so-called q-logarithm: $\ln_q(x) = (x^{1-q} - 1)/(1-q)$. Our algorithm estimates for each $\delta_q = 0,01$ step the linear adjustment on the graph under scrutiny (in this case the $\ln_q(p(z_i)) \, vs \; z_i^2$ graph) by evaluating the associated correlation coefficient *(CC)*, while the best linear fit is considered to be the one maximizing the correlation coefficient. The obtained $q_{stat}$, corresponding to the best linear adjustment is then being used to compute the following equation:

$$G_q(\beta, z) = \frac{\sqrt{\beta}}{C_q} e_q^{-\beta z^2}$$ (2.63)

where $C_q = \sqrt{\pi} \cdot \Gamma(\frac{3-q}{2(q-1)}) / \sqrt{q-1} \cdot \Gamma(\frac{1}{q-1})$, $1 < q < 3$ for different $\beta$-values. Moreover, we select the $\beta$-value minimizing the $\sum_i [G_{q_{stat}}(\beta, z_i) - p(z_i)]^2$, as proposed again in [79].

Finally the $q_{rel}$ index is given by the relation: $q_{rel} = (s-1)/s$, where $s$ is the slope of the log-log plotting of the autocorrelation function of the sunspot index, according to the relaxation process:

$$\log C(\tau) = a + s \log(\tau), \qquad (2.64)$$

where $C(\tau) = < [Z(t_i + \tau) - < Z(t_i) >] - [Z(t_i) - < Z(t_i) >] / < Z(t_i) - < Z(t_i) >]^2 >$. $\qquad (2.65)$

Instead of $C(\tau)$ we can use also the relaxation of mutual information $I(\tau)$ given in Fraser and Swinney [80] by the relation:

$$I(\tau) = -\sum_{x(i)} P(x(i)) \log_2 P(x(i)) - \sum_{x(i-\tau)} P(x(i-\tau)) \log_2 P(x(i-\tau)) + \sum_{x(i)} \sum_{x(i-\tau)} P(x(i), x(i-\tau)) \log_2 P(x(i), x(i-\tau)) \qquad (2.66)$$

## 2.2.4 Structure Functions Scaling Exponent Spectrum

The $P_{th}$ order structure function of the observed signal timeseries is given by the relation:

$$S^p(\tau) \equiv < |V(t+\tau) - V(t)|^p > = \tau^{J(p)}, \qquad (2.67)$$

where in our analysis $p$ may range from 1-20. When the experimental signal can be related to a characteristic $(V_{ts})$ flow speed phenomenon, then the length time lag $\tau$ corresponds to the flow speed of eddies of size $l = \tau V_{fs}$. In the case of solar plasma the magnetic field frozen condition can be used for extracting also spatial characteristics of the turbulent solar magnetic field dissipations. For turbulence with Gaussian statistics a universal scaling law was established at first by Kolmogorov [68], according to the relation:

$$S_p(\tau) \sim \tau^{J(p)} \qquad (2.68)$$

where $J(p) = p/3$. According to previous description of intermittent turbulence, the turbulent eddies are not homogeneous but they occupy at every scale only a fraction with dimension $D_F$, of the volume element given by the factor $2^{(D_F-3)}$. In this case as intermittency prevails there is a departure from the simple low $J = p/3$. While, the scaling exponent is given [71,81] by:

$$J(p) = 3 - D_F + p(D_F - 2)/3. \qquad (2.69)$$

In this study the scaling exponent spectrum $S(p)$ is estimated experimentally by log-log plotting of the structure function $S^p(\tau), p = 2-16$, as well as by using the Tsallis statistical theory as it was described in section 2.1.

## 2.2.5 Correlation Dimension

In order to provide information for the dynamical degrees of freedom of the dynamics underlying the experimental time series we estimate the correlation dimension $(D)$ of the dynamical trajectory of the system in the reconstructed state space of embedding dimension $m$:

$$\mathbf{x}(t_i) = \left[ x(t_i), x(t_i + \tau), ..., x(t_i + (m-1)\tau) \right] \qquad (2.70)$$

where $x(t_i)$ is the observed time series, as the saturation value of the slopes $D_m$ of the correlation integrals according to the relation:

$$D = \lim_{\substack{r \to 0 \\ m \to \infty}} D_m \qquad (2.71)$$

where $D_m = \lim_{r \to 0} \dfrac{d \ln C_m(\mathbf{r}, m)}{d \ln(r)}$ and

$$C_m(\mathbf{r}, m) = \frac{2}{N(N-1)} \sum_{i=1}^{N} \sum_{j=1}^{N} \Theta(r - \| \mathbf{x}(i) - \mathbf{x}(j) \|) \qquad (2.72)$$

is the correlation integral of the trajectory estimated in the *m*-dimensional reconstructed state space. The low value saturation of the slopes of the correlation integrals is related to the number (*d*) of fundamental degrees of freedom of the internal dynamics. For the estimation of the correlation integral we used the method of Theiler [82] in order to exclude time correlated states in the correlation integral estimation, thus discriminating between the dynamical character of the correlation integral scaling and the low value saturation of slopes characterizing self-affinity (or crinkliness) of trajectories in a Brownian process. When the dynamics possesses a finite (small) number of degrees of freedom, we can observe saturation to low values $D$ of the slopes $D_m$ obtained in (2.65) for a sufficiently large embedding *m*. The dimension of the attractor of the dynamics is then at least the smallest integer $(D_0)$ larger than $D$ or at most $2D_0 + 1$, according to Taken's theorem [74].

### 2.2.6 Lyapunov Exponent Spectrum

The spectrum of the Lyapunov exponents ($\lambda_j$) can be found by following the evolution of small perturbations of the dynamical orbit in the reconstructed state space according to the relations:

$$\frac{d\vec{W}}{dt} = \bar{A}\vec{W} \qquad (2.73)$$

and

$$\vec{W}(t) = \sum C_j \hat{e}_j \exp(\lambda_j t), \qquad (2.74)$$

where $C_j$ are coefficients determined by the initial conditions and $\hat{e}_j$ are the eigenvectors of the evolution matrix ($\bar{A}$) corresponding to different eigenvalues ($\lambda_j$). For the estimation of the entire Lyapunov spectrum we follow Sano and Sawada [83] correspondingly.

### 2.2.7 Surrogate Data

According to Theiler the method of surrogate data is used to distinguish between linearity and nonlinearity as well as between chaoticity and pure stochasticity, since a linear stochastic signal can mimic a nonlinear chaotic process after a static nonlinear distortion [34,35]. Surrogate data are constructed according to Schreiber & Schmitz [84]to mimic the original data, regarding their autocorrelation and amplitude distribution. In particular, the procedure starts with a white noise signal, in which the Fourier amplitudes are replaced by the corresponding amplitudes of the original data. In the second step, the rank order of the derived stochastic signal is used to reorder the original time series. By doing this, the amplitude distribution is preserved, but the matching of the two power spectra achieved at the first step is altered. The two steps are subsequently repeated several times until the change in the matching of the power spectra is sufficiently small. Surrogate data thus provide the most general type of nonlinear stochastic (white noise) signals that can approach the geometrical or dynamical characteristics of the original data. They can be used for the rejection of every null hypothesis that identifies the observed low dimensional chaotic profile as a purely non-chaotic stochastic linear process. For an extensive description of the nonlinear analysis algorithm [37,38].

In order to distinguish a nonlinear deterministic process from a linear stochastic one, we use as discriminating statistic a quantity Q derived from a method sensitive to nonlinearity, for example the correlation dimension, the maximum Lyapunov exponent, the mutual information etc. The discriminating statistic Q is then calculated for the original and the surrogate data and the null hypothesis is verified or rejected depending on the ''number of sigmas''.

## 3. Results of Data Analysis

Sunspots are temporary phenomena on the photospheric surface of the convection zone of the sun. The sunspots are the visible counterparts of magnetic flux – tubes rising in the sun's convective zone. The intense magnetic activity at the sunspot regions inhibits plasma convection forming areas of reduced surface temperature. As they move across the surface of the sun, the sunspots expand and contrast in a diameter of $\sim 8.000$ km. The Wolf number, known as the international sunspot number measures the number of sunspots and group of sunspots on the surface of the sun computed by the formula: $R = k(10g + s)$ where: $s$ is the number of individual spots, $g$ is the number of sunspot groups and $k$ is a factor that varies with location known as the observatory factor.

In this section we present results concerning the analysis of data included in the sunspot index by following physical and the methodology included in the previous section of this study.

### 3.1 Time series and Flatness Coefficient F.

Fig.1a represents the Sunspot Index Time series that was constructed concerning the period of 184 years. As it can be seen from this figure the signal has a strong periodic component, almost every 4300 days that corresponds to the well known phenomenon of Solar Cycle. Fig1b presents the flatness coefficient F estimated for the sunspot data during the same period. The F values reveal continues variation of the sunspot statistics between Gaussian profile ($F \sim 3$) to strong non-Gaussian profile ($F \sim 4 - 12$). Fig.1c presents the first ($V_1$) SVD component and Fig.1d the sum $V_{2-10} = \sum_{i=2}^{10} V_i$ of the next SVD components estimated for the sunspot index time series. The estimation of the flatness coefficient $F(V_1)$ and $F(V_{2-10})$ is shown in Fig.1e and Fig.1f, correspondingly. As we can notice in these figures the statistics of the $V_1$ component is clearly discriminated from the statistics of the $V_{2-10}$ component, as the $F(V_1)$ flatness coefficient obtained almost every where low values ($\sim 3 - 4$) while the $F(V_{2-10})$ obtained higher values($\sim 4 - 9$). The low values ($\sim 3 - 4$) of the $F(V_1)$ coefficient indicates for the solar activity a near Gaussian dynamical process, underlying the $V_1$ SVD component. Oppositely, the high values of the $F(V_{2-10})$ coefficient indicate strongly non-Gaussian solar dynamical process underlying the $V_{2-10}$ SVD component. According to the physical meaning of the SVD analysis (section 2.2.1), the $V_1$ SVD component can be related to the solar dynamics extended at large spatio-temporal regions of the photospheric system, while the $V_{2-10}$ SVD component of the sunspot index is related to the spatio-temporal confined photospheric dynamics.

Figure 1

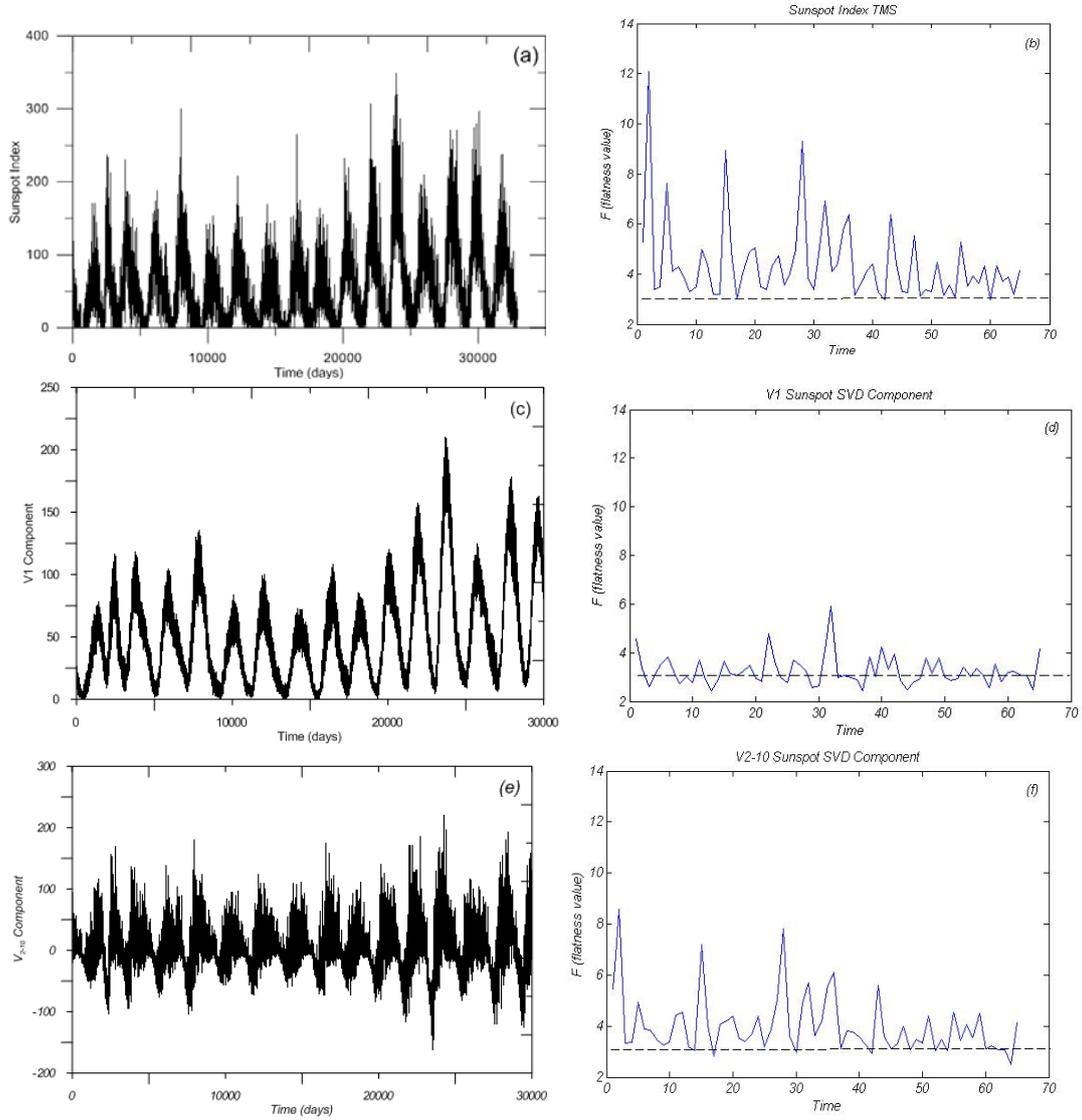

**Figure 1**: **(a)** *Time series of Sunspot Index concerning the period of 184 years.* **(b)** *The coefficient F estimated for Sunspot Index TMS* **(c)** *The SVD $V_1$ component of Sunspot Index tms.* **(d)** *The coefficient F estimated for the SVD $V_1$ component of Sunspot Index tms.* **(e)** *The SVD $V_{2-10}$ component of Sunspot Index tms.* **(f)** *The coefficient F estimated for the SVD $V_{2-10}$ component of Sunspot Index tms.*

### 3.2 The Tsallis q-statistics

In this section we present results concerning the computation of the Tsallis q-triplet, including the three-index set ($q_{stat}, q_{sen}, q_{rel}$) estimated for the original sunspot index timeseries, as well as for its $V_1$ and $V_{2-10}$ SVD components (presented in Fig.1[a-c,e].

### 3.2.2 Determination of $q_{stat}$ index of the q-statistics.

In Fig.2a we present (by open circles) the experimental probability distribution function (PDF) p(z) vs. z, where z corresponds to the $Z_{n+1} - Z_n, (n = 1, 2, ..., N)$ timeseries difference values. In Fig.2b we present the best linear correlation between $\ln_q[p(z)]$ and $z^2$. The best fitting was found for the value of $q_{stat} = 1.53 \pm 0.04$. This value was used to estimate the q-Gaussian distribution presented in Fig.2a by the solid black line. Fig.2[c,d] and Fig.2[e,f] are similar to Fig.2[a,b] but for the $V_1$ and $V_{2-10}$ SVD components correspondingly. Now the $q_{stat}$ values were for the SVD components estimated to be: $q_{stat}(V_1) = 1.40 \pm 0.08$ and $q_{stat}(V_{2-10}) = 2.12 \pm 0.20$. As we can observed from these results the following relation is satisfied:
$1 < q_{stat}(V_1) < q_{stat}(orig) < q_{stat}(V_{2-10})$, where $q_{stat}(orig)$ corresponds to the original sunspot index timeseries.

### 3.2.2 Determination of $q_{rel}$ index of the q-statistics.
### a) Relaxation of autocorrelation functions

Fig.3 presents the best log plot fitting of the autocorrelation function $C(\tau)$ estimated for the original sunspot index signal (Fig.3a) its $V_1$ SVD component (Fig.3b), as well as its $V_{2-10}$ SVD component (Fig.3c). The three $q_{rel}$ values were found to satisfy the relation: $1 < q_{rel}^c(V_{2-10}) < q_{rel}^c(orig) < q_{rel}^c(V_1)$ as:
$q_{rel}^c(V_{2-10}) = 4.115 \pm 0.134$, $q_{rel}^c(V_{orig}) = 5.672 \pm 0.127$, $q_{rel}^c(V_1) = 29.571 \pm 0.794$

### b) Relaxation of Mutual Information
Fig.3[b,d,f] is similar with Fig.3[a,c,e] but it corresponds to the relaxation time of the mutual information $I(\tau)$. For the top of the bottom we see the log-log plot of $I(\tau)$ for the sunspot index timeseries, its $V_1$ SVD component and its $V_{2-10}$ SVD component. The best log-log (linear) fitting showed the values:
$q_{rel}^I(V_{2-10}) = 2.426 \pm 0.054$, $q_{rel}^I(V_{orig}) = 2.522 \pm 0.044$, $q_{rel}^I(V_1) = 5.255 \pm 0.308$. Among the three values it is satisfied the following relation: $1 < q_{rel}^I(V_{2-10}) < q_{rel}^I(orig) < q_{rel}^I(V_1)$. Also comparing the $q_{rel}$ indices, as they were estimated for the autocorrelation function and the mutual information function, it was found the following relation: $q_{rel}^I < q_{rel}^c$ for all the cases. The last result is explained by the fact that the mutual information includes nonlinear characteristics of the underlying dynamics in contrast to the autocorrelation function which is a linear statistical index.

Figure 2

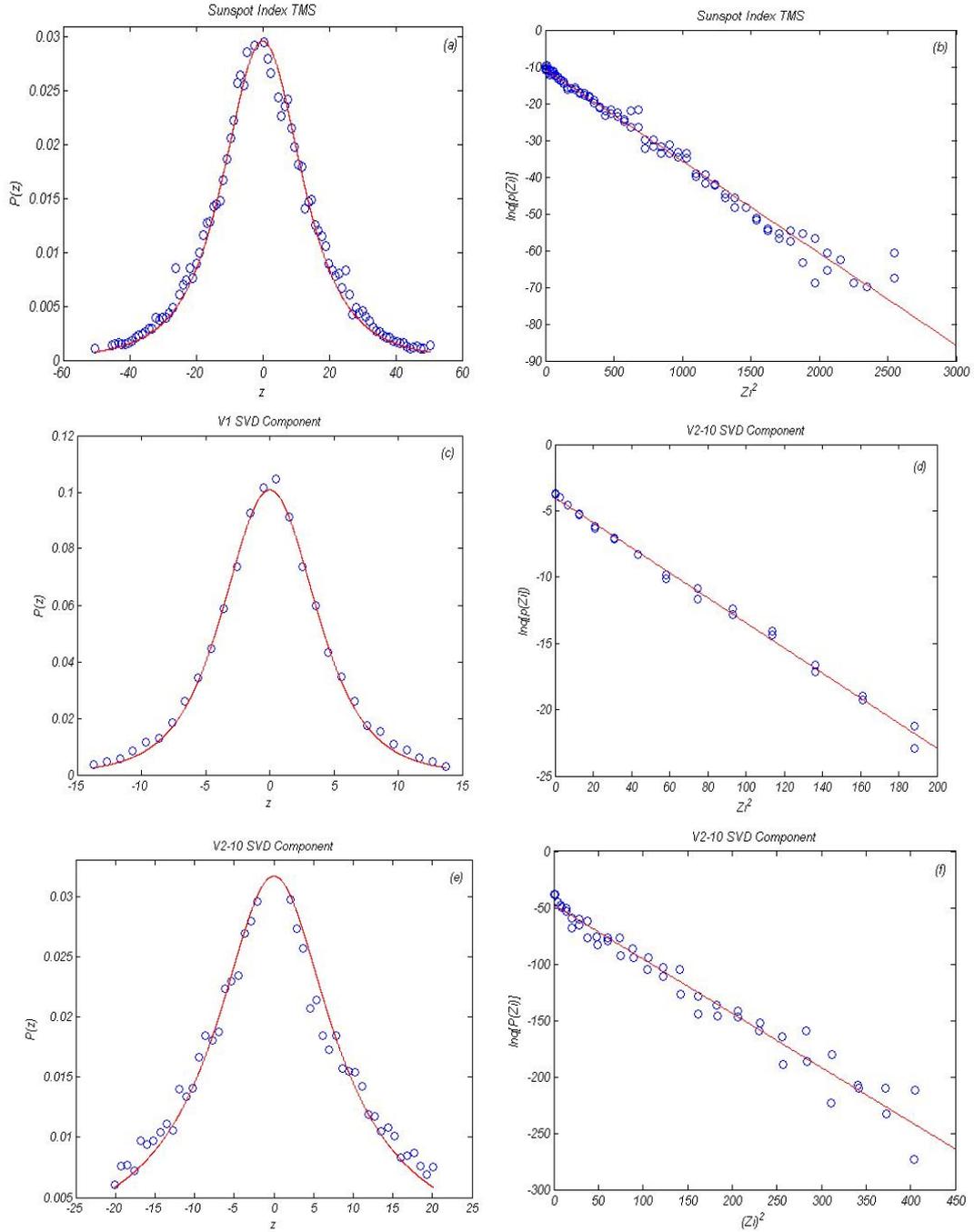

***Figure 2:*** ***(a)*** *PDF P($z_i$) vs. $z_i$ q Guassian function that fits P($z_i$) for the Sunspot Index timeseries* ***(b)*** *Linear Correlation between $\ln_q p(z_i)$ and $(z_i)^2$ where q = 1.53 ± 0.04 for the Sunspot Index timeseries* ***(c)*** *PDF P($z_i$) vs. $z_i$ q Gaussian function that fits P($z_i$) for the $V_1$ SVD component* ***(d)*** *Linear Correlation between $\ln_q p(z_i)$ and $(z_i)^2$ where q = 1.40 ± 0.08 for the $V_1$ SVD component* ***(e)*** *PDF P($z_i$) vs. $z_i$ q Gaussian function that fits P($z_i$) for the $V_{2-10}$ SVD component* ***(f)*** *Linear Correlation between $\ln_q p(z_i)$ and $(z_i)^2$ where q = 2.12 ± 0.20 for the $V_{2-10}$ SVD component*

Figure 3

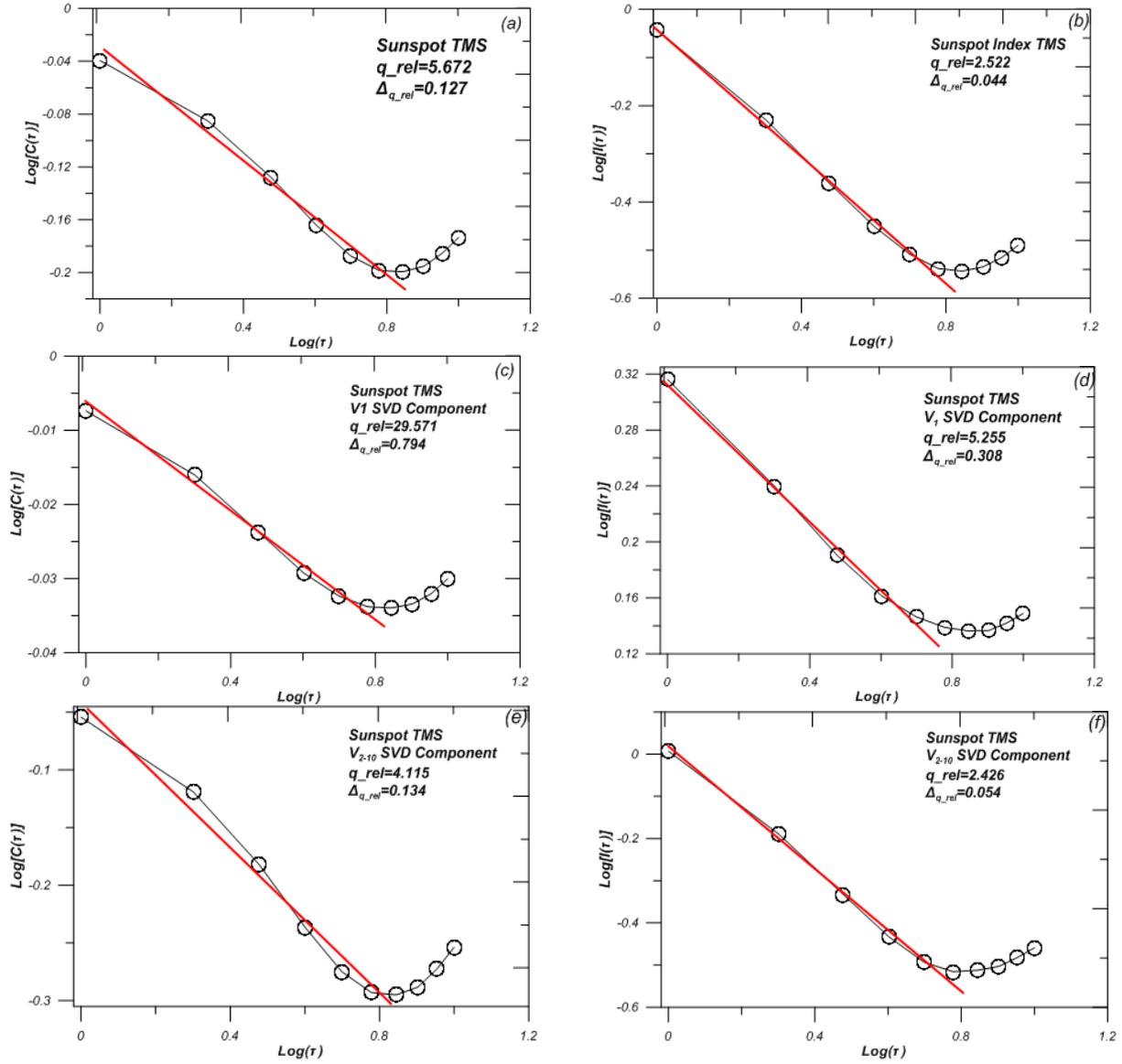

***Figure 3***: (***a***) *Log – log plot of the self-correlation coefficient C(τ) vs. time delay τ for the Sunspot Index time series. We obtain the best fit with qrel=5.672±0.127 (**b**)Log – log plot of the mutual information I(τ) vs. time delay τ. for the Sunspot Index time series. We obtain the best fit with qrel=2.522±0.044 (**c**) Log – log plot of the self-correlation coefficient C(τ) vs. time delay τ for the $V_1$ SVD component. We obtain the best fit with qrel=29.571±0.794 (**d**)Log – log plot of the mutual information I(τ) vs. time delay τ for the $V_1$ SVD component. We obtain the best fit with qrel=5.255±0.308. (**e**) Log – log plot of the self-correlation coefficient C(τ) vs. time delay τ for the $V_{2-10}$ SVD component. We obtain the best fit with qrel=4.115±0.134 (**f**)Log – log plot of the mutual information I(τ) vs. time delay τ. for the $V_{2-10}$ SVD component. We obtain the best fit with qrel=2.426±0.054.*

### 3.2.3 Determination of $q_{sen}$ index of the q-statistics.

Fig.4 presents the estimation of the generalized dimension $D_q$ and their corresponding multifractal (or singularity) spectrum $f(\alpha)$, from which the $q_{sen}$ index was estimated by using the relation $1/(1-q_{sens}) = 1/a_{min} - 1/a_{max}$ for the original sunspot index (Fig.4 [a-b], as well as its $V_1$ (Fig.4[c-d]) and $V_{2-10}$ (Fig.4[e-f]) SVD components. The three index $q_{sen}$ values were found to satisfy the relation: $1 < q_{sen}(V_1) < q_{sen}(orig) < q_{sen}(V_{2-10})$, where: $q_{sen}(V_1) = 0.055 \pm 0.009$, $q_{sen}(V^{orig}) = 0.368 \pm 0.005$ and $q_{sen}(V_{2-10}) = 0.407 \pm 0.029$.

In Fig.4[a,c,e] the experimentally estimated spectrum function $f(\alpha)$ is compared with a polynomial of sixth order (solid line) as well as by the theoretically estimated function $f(\alpha)$ (dashed line), by using the Tsallis $q-$entropy principle according to the relation (2.25). As we can observe the theoretical estimation is faithful with high precision on the left part of the experimental function $f(a)$. However, the fit of theoretical and experimental data are less faithful for the right data, especially for the original sunspot timeseries (Fig.4a) and its SVD components $V_{2-10}$ (Fig.4e). Finally, the coincidence of theoretically and experimentally data is excellent for the $V_1$ SVD component (Fig.4c).

Similar comparison of the theoretical prediction and the experimental estimation of the generalized dimensions function $D(q)$ is shown in Fig.4(b,d,f). In these figures the solid brown line correspond to the $p-$model prediction according to [65], while the solid red line correspond to the $D(q)$ function estimation according to Tsallis theory [67,69].

The correlation coefficient of the fitting was found higher than 0.9 for all cases. Also, the parameter $p$ of the $p-$model, as evaluated from the nonlinear best fitting of $D_q$ was found as follows: $p(orig) = 0.772$, $p(V_1) = 0.692$, $p(V_{2-10}) = 0.797$. These values are different from the value $p = 0.5$ corresponding to the Gaussian turbulence, satisfying also the relation: $p(V_1) < p(orig) < p(V_{2-10})$. These results indicates for the turbulence cascade of solar plasma, partial mixing and asymmetric (intermittent) fragmentation process of the energy dissipation.

It is interesting here to notice also the relation between the $\Delta\alpha_{min,max}$ and $\Delta D_{-\infty,+\infty}$ values where $\Delta\alpha_{min,max} = a_{max} - a_{min}$ and $\Delta D_{-\infty,+\infty} = D_{q-\infty} - D_{q+\infty}$, which was found to satisfy the following ordering relation:

$\Delta\alpha(V_1) = 1.113 < \Delta\alpha(orig) = 1.752 < \Delta\alpha(V_{2-10}) = 1.940$,
$\Delta Dq(V_1) = 0.234 < \Delta Dq(orig) = 1.367 < \Delta Dq(V_{2-10}) = 1.517$

The $\Delta D_{-\infty,+\infty}$ and $\Delta\alpha_{min,max}$ values were estimated also separately for the first seven SVD components. The $\Delta D_{-\infty,+\infty}(V_i)$ values vs. $V_i, (i = 1,...,7)$ are showed in Fig.5a and the $\Delta\alpha_{min,max}(V_i)$ vs. $V_i, (i = 1,...,7)$ are showed in Fig.5b. The spectra of $\Delta D_q$ and $\Delta D_a$ shown in these figures (Fig.5[a,b]) reveal positive and increasing profile as we pass from the first to the last SVD component. This clearly indicated intermittent solar turbulent process underlying all the SVD components of the sunspot index.

Figure 4

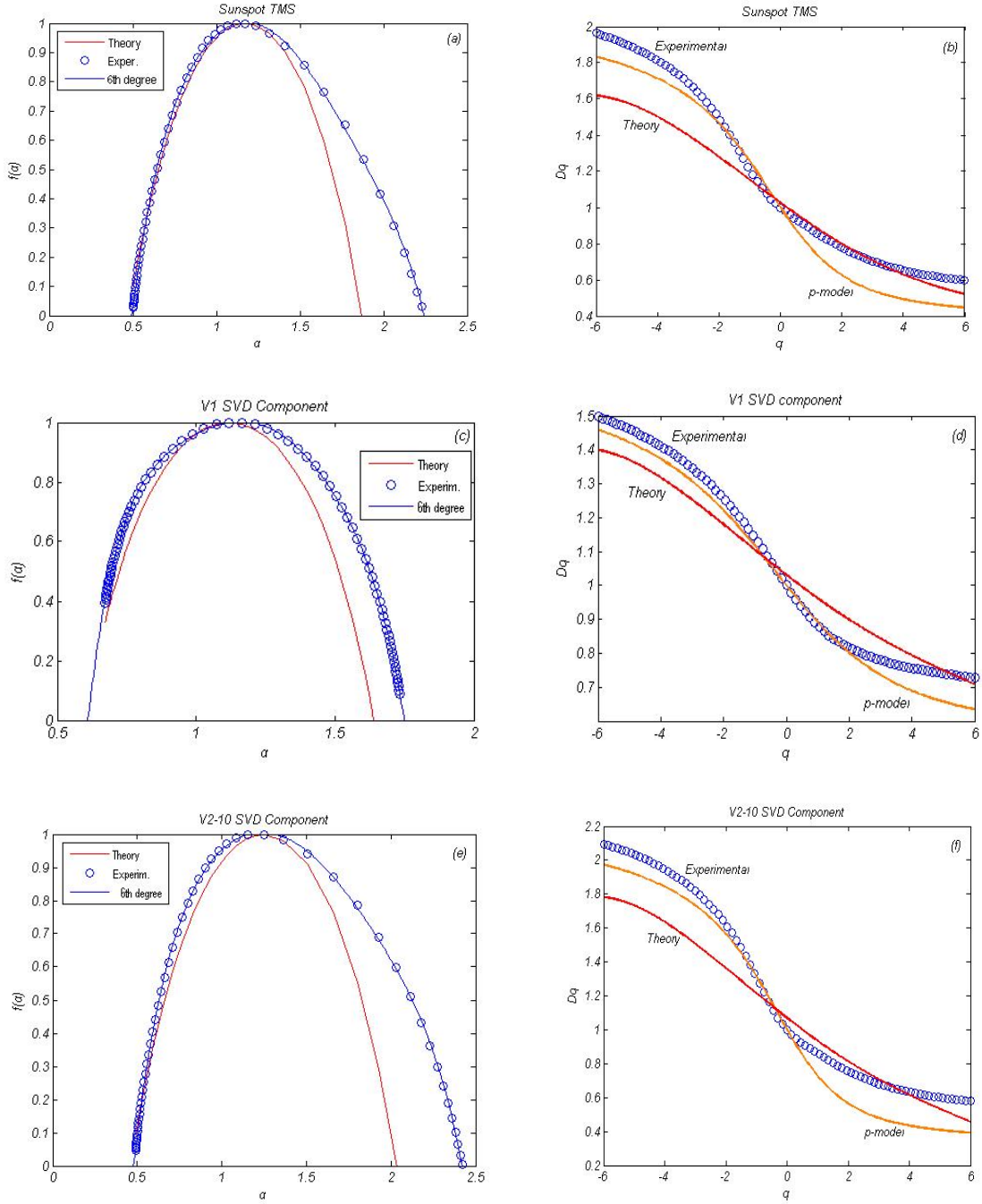

*Fig4: (a) Multifractal spectrum of Sunspot timeseries. $a_{min} = 0.493 \pm 0.003$. and $a_{max} = 2.245 \pm 0.031$. With solid line a sixth-tenth degree polynomial. We calculate the $q\_sen=0.368\pm0,005$. (b) $D(q)$ vs. $q$ of Sunspot timeseries. (c) Multifractal spectrum of $V_1$ SVD component. $a_{min} = 0.613 \pm 0.004$. and $a_{max} = 1.746 \pm 0.003$. We calculate the $q\_sen=0.055\pm0,009$. With solid line a sixth-tenth degree polynomial (d) $D(q)$ vs. $q$ of $V_1$ SVD component. (e) Multifractal spectrum of $V_{2-10}$ SVD component. $a_{min} = 0.476 \pm 0.019$. and $a_{max} = 2.416 \pm 0.002$. We calculate the $q\_sen=0.407\pm0.029$ . With solid line a sixth-tenth degree polynomial (f) $D(q)$ vs. $q$ of $V_{2-10}$ SVD component.*

Figure 5

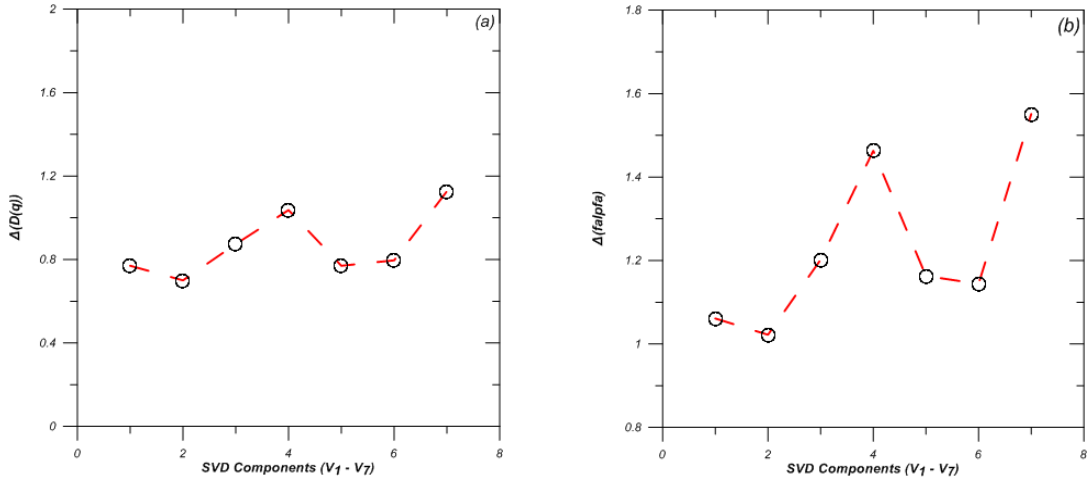

*Figure 5: (a) The differences Δ(Dq(Vi)) versus the Vi SVD components of Sunspot timeseries for i=1,2,...,7. (b) The differences Δ(Da(Vi)) versus the Vi SVD components of Sunspot timeseries for i=1,2,...,7.*

### 3.3 Determination of Structure Function spectrum
### 3.3.1 Intermittent Solar Turbulence

Fig.6[a-c] shows the structure function $S(p)$ plotted versus time lag ($\tau$) estimated for the original sunspot index signal (Fig.6a) as well as for its SVD components $V_1$ (Fig.6b) and $V_{2-10}$ (Fig.6c). At low values of time lag ($\tau$), we can observe scaling profile for all the cases of the original timeseries and its SVD components for every p value. Fig.6[d-f] presents the best linear fitting of the scaling regions in the lag time interval $\Delta \log(\tau) = 0 - 1.6$. Fig.7a shows the exponent $J(p)$ spectrum of structure function vs. the *pth* order estimated separately for the first seven SVD components $V_i, (i = 1,...,7)$. In the same figure we present the exponent $J(p)$ spectrum of the structure function $S(p)$ according to the original theory of Kolmogorov [68] for the fully developed turbulence, of Gaussian turbulence known as $p/3$ theory. Here we can observe clear discrimination from the Gaussian turbulence K41 theory for all the SVD components. Moreover we observe significant dispersion of the $J(p)$ spectrum from the first to the seventh SVD components, especially for the higher orders ($p > 10$).

Fig.7c is similar to Fig.7a but for the $V_1$ SVD component and the summarized $V_{2-10}$ SVD component. Also, here we notice significant discrimination between the $V_1$ and $V_{2-10}$ SVD components while both of them reveal *pth* order structure functions with their own index different from the $p/3$ values given by the K41 theory. In order to study furthermore the departure from the Gaussian behavior of the turbulence underlying the sunspot index signal, we

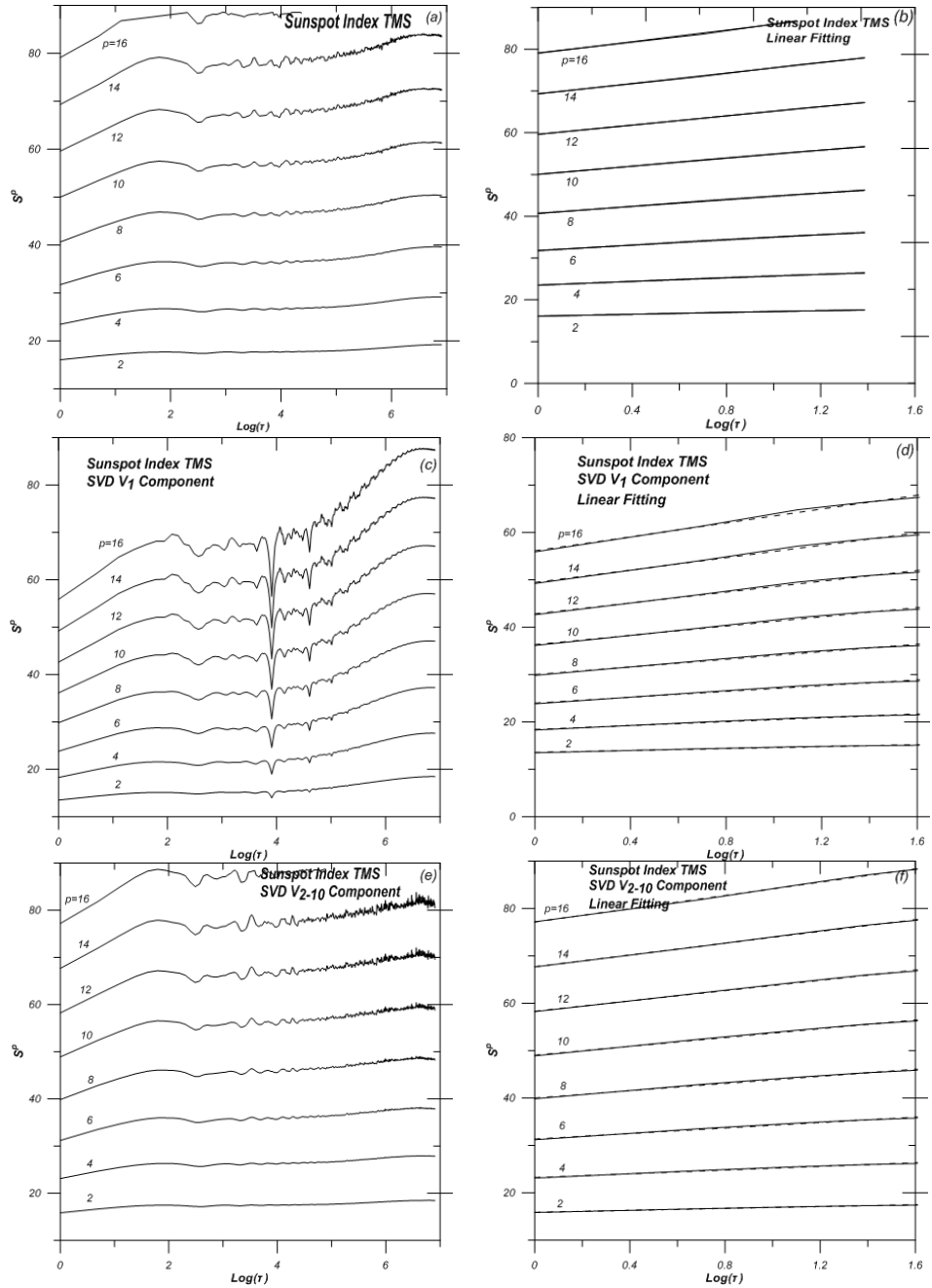

Figure 6: (a) The log – log plot of structure function $S^p$ of the Sunspot index timeseries vs. time lag $\tau$ for various values of the order parameter p. (b)  The first linear scaling of the log – log plot. (c) The log – log plot of structure function $S^p$ of the $V_1$ SVD component vs. time lag $\tau$ for various values of the order parameter p. (d)  The first linear scaling of the log – log plot. (e) The log – log plot of structure function $S^p$ of the $V_{2-10}$ SVD component vs. time lag $\tau$ for various values of the order parameter p. (f)  The first linear scaling of the log – log plot.

Figure 7

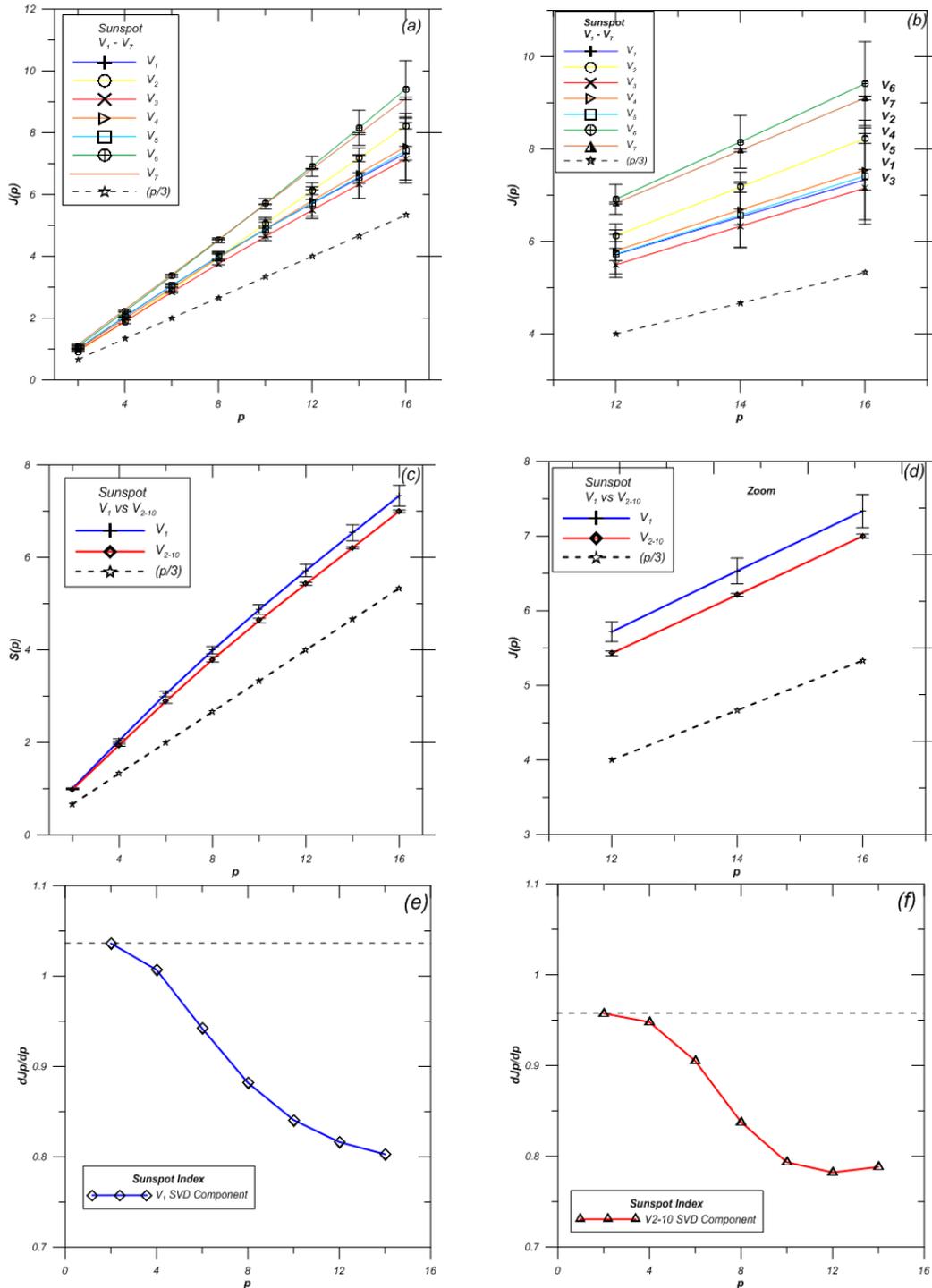

*Figure 7: (a) The scaling exponent J(p) versus p of the independent $V_1$-$V_7$ SVD components of the Sunspot timeseries and compared with the Kolmogorov p/3 prediction (dashed line) (b) The zoom in the area of p=12-16. The scaling exponent J(p) versus p of the $V_1$, $V_{2-10}$ SVD components of the Sunspot timeseries and compared with the Kolmogorov p/3 prediction (dashed line) (b) The zoom in the area of p=12-16. (e)The h(p)=dJp/dp function versus p for the $V_1$ SVD component of Sunspot timeseries. (f)The h(p)=dJp/dp function versus p for the $V_{2-10}$ SVD component of Sunspot timeseries.*

Figure 8

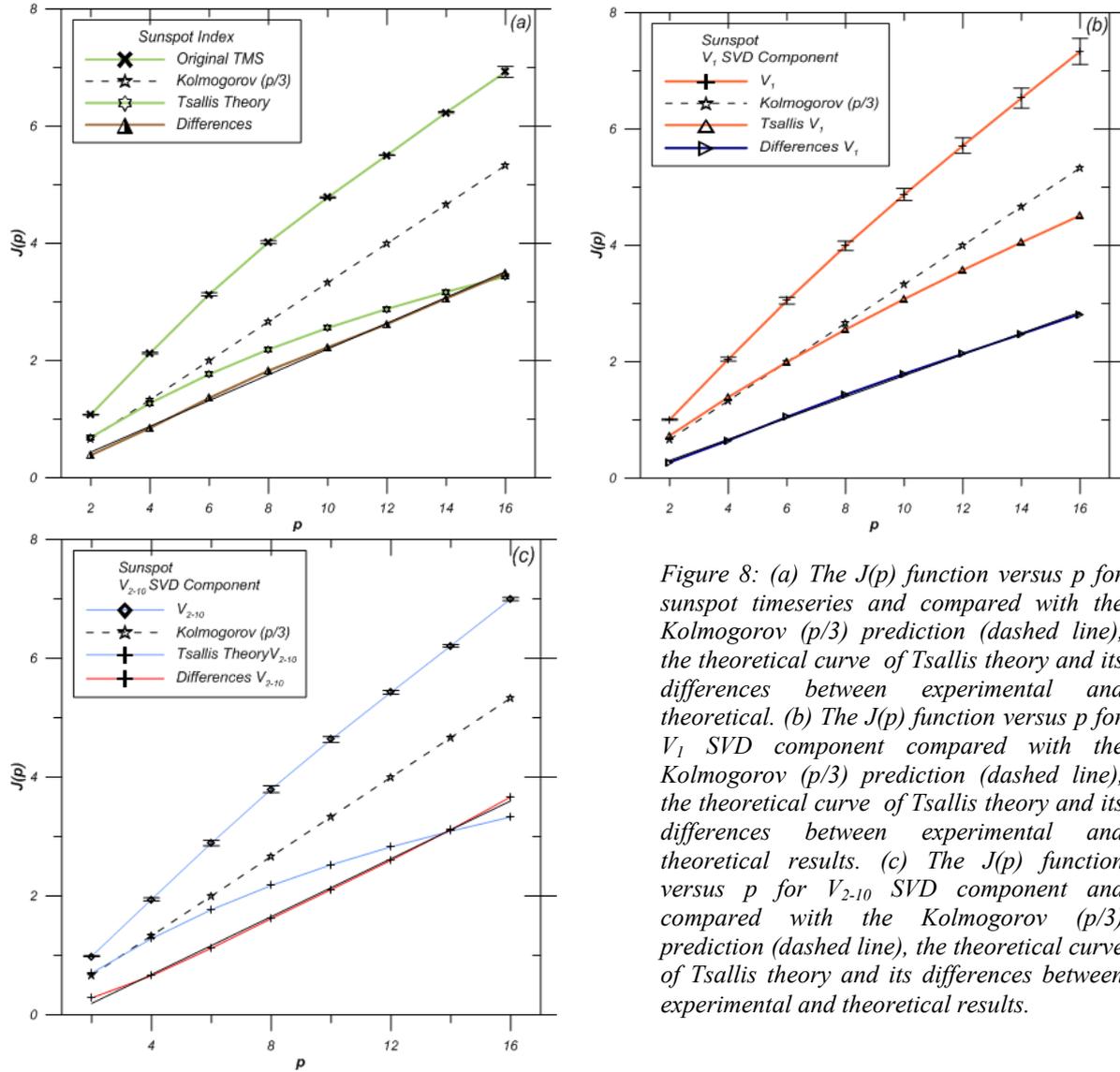

*Figure 8: (a) The J(p) function versus p for sunspot timeseries and compared with the Kolmogorov (p/3) prediction (dashed line), the theoretical curve of Tsallis theory and its differences between experimental and theoretical. (b) The J(p) function versus p for $V_1$ SVD component compared with the Kolmogorov (p/3) prediction (dashed line), the theoretical curve of Tsallis theory and its differences between experimental and theoretical results. (c) The J(p) function versus p for $V_{2-10}$ SVD component and compared with the Kolmogorov (p/3) prediction (dashed line), the theoretical curve of Tsallis theory and its differences between experimental and theoretical results.*

.

present in Fig.7[c-d] the derivatives $h(p) = dJ(p)/dp$ of the structure functions estimated for the spectra $S(p)$ of the $V_1$ and $V_{2-10}$ SVD components. For both cases the non-linearity of the functions $S_{V_1}(p), S_{V_{2-10}}(p)$ is apparent as their derivatives are strongly dependent upon the $p$ values.

### 3.3.2` Comparison of solar turbulence with non-extensive q-statistics.

In this section we present interesting results concerning the comparison of the structure function experimental estimation with the theoretical predictions according to Arimitsu and Arimitsu [67] by using the Tsallis non-extensive statistical theory, as it was presented at the description of section 2. Fig.8a presents the structure function $J(p)$ vs. the order parameter $p$, estimated for the sunspot index timeseries (black line) the K41 theory (dashed line), the theoretically predicted values of structure function (red line) by using q-statistics theory, as well as the differences $\Delta J(p)$ vs. $p$, between the experimentally and theoretically produced values. Fig.8b is similar to Fig.8a but corresponding to the $V_1$ and $V_{2-10}$ SVD components of the original sunspot index timeseries. For all the cases presented in Fig.8[a-b] the theoretically estimated $J(p)$ values (red lines) correspond to the HD turbulent dissipation of the solar plasma, revealing values lower than the K41 prediction in accordance with the HD intermittent turbulence.

According to previous theoretical description (section 2) the experimentally produced structure function spectrum of the original signal and its SVD components is caused by including kinetic and magnetic dissipation simultaneously [71] the MHD solar turbulence. In this case the solar magnetic field dissipation makes the structure function spectrum to obtain values much higher than the values of the corresponding HD intermittent turbulence according to the relation (2.40). As we can observe in Fig.8 [a-b] the best fitting to the differences $\Delta J(p)$ shows for all cases (the sunspot index and its $V_1, V_{2-10}$ SVD components) the existence of the linear relation: $\Delta J(p) \approx \alpha(p-b)$.

The values of the $\alpha, b$ parameters were estimated as follow:

$\alpha_{orig} = 0.219, b_{orig} = 0.0004, \alpha_{V_1} = 0.182, b_{V_1} = -0.057, \alpha_{V_{2-10}} = 0.243, b_{V_{2-10}} = -0.293$

### 3.5 Determination of Correlation Dimension

### 3.5.1 Correlation dimension of the Sunspot timeseries

Fig.9a shows the slopes of the correlation integrals vs. $\log r$ estimated for the sunspot index timeseries for embedding dimension $m = 6 - 10$. As we notice here there is no tendency for low value saturation of the slopes which increase continuously as the embedding dimension ($m$) increases. The comparison of the slopes with surrogate data, showed in Fig.9[b-c]. Fig.9b presents the slopes of the original signal at embedding dimension $m = 7$ (red line), as well as the slopes estimated for a group of corresponding surrogate data according to section (2.2.7). The significance of the statistics was shown in Fig.9d. As the significance of the discriminating statistics remains much lower than two sigmas, we cannot reject the null hypothesis of a high dimensional Gaussian and linear dynamics underlying the Sunspot index of the solar activity.

Fig.10 shows results concerning the estimation correlations dimensions of the $V_1$ and $V_{2-10}$ SVD components of the original sunspot index signal.

### 3.5.1 Correlation dimension of the $V_1$ SVD component

The slopes of the correlation integrals estimated for the $V_1$ SVD component and its corresponding surrogate data are presented in Fig.10a and Fig.10b. The slope profiles of the $V_1$ SVD component reveal a tendency for low value saturation at values lower than 6-7. However the corresponding slope profiles of the surrogate data are similar with the slopes profile of the $V_1$ SVD component. Fig.10c shows the slopes of the $V_1$ SVD component and a group of corresponding surrogate data at the embedding dimension $m = 10$, while in Fig.10d we present the significance of the statistics. As the significance remains at low values ($< 2 sigmas$) for small values of $\log r$ it is not possible the rejection of null hypothesis.

Figure 9

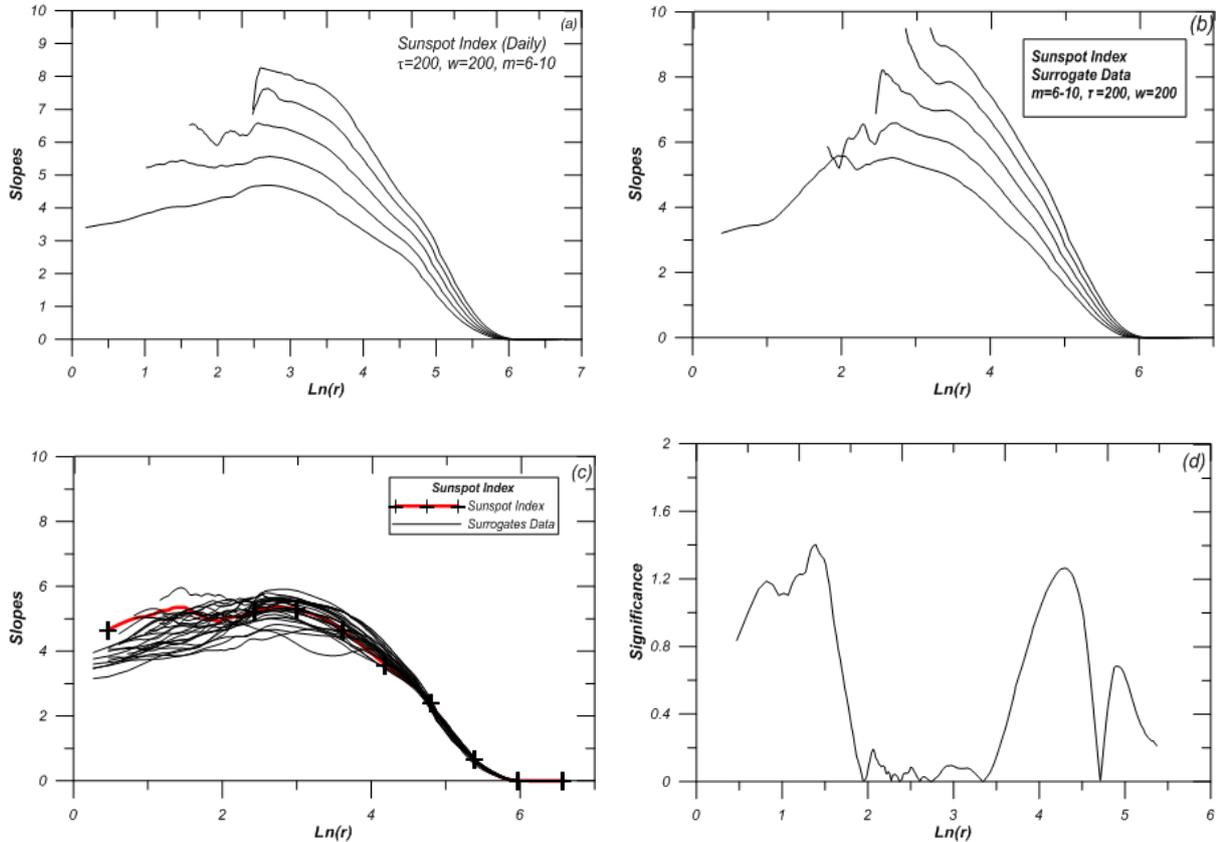

***Figure 9:*** *(**a**) Slopes D of the correlation integrals estimated for the Sunspot timeseries. (**b**) Slopes D of the correlation integrals estimated for the Surrogate timeseries. (**c**) Slopes of the correlation integrals of the Sunspot time series and its thirty (30) surrogate estimated for delay time τ=200 and for embedding dimension m=7, as a function of Ln(r). (**d**) The Significance of the Statistics for the sunspot timeseries and its 30 Surrogates.*

### 3.5.2 Correlation dimension of the $V_{2-10}$ SVD component

Fig.10[e-h] is similar to Fig.10[a-d] but for the $V_{2-10}$ SVD component. The slope profiles of this signal reveal clearly low value saturation at value lower ~ 6 (Fig.10e) with possibility of strong discrimination from the corresponding surrogate data Fig.10[f,g]. The significance of the discriminating statistics is much higher than two sigmas (Fig.10h.). This permits the rejection of the null hypothesis with confidence $> 99\%$.

Figure 10

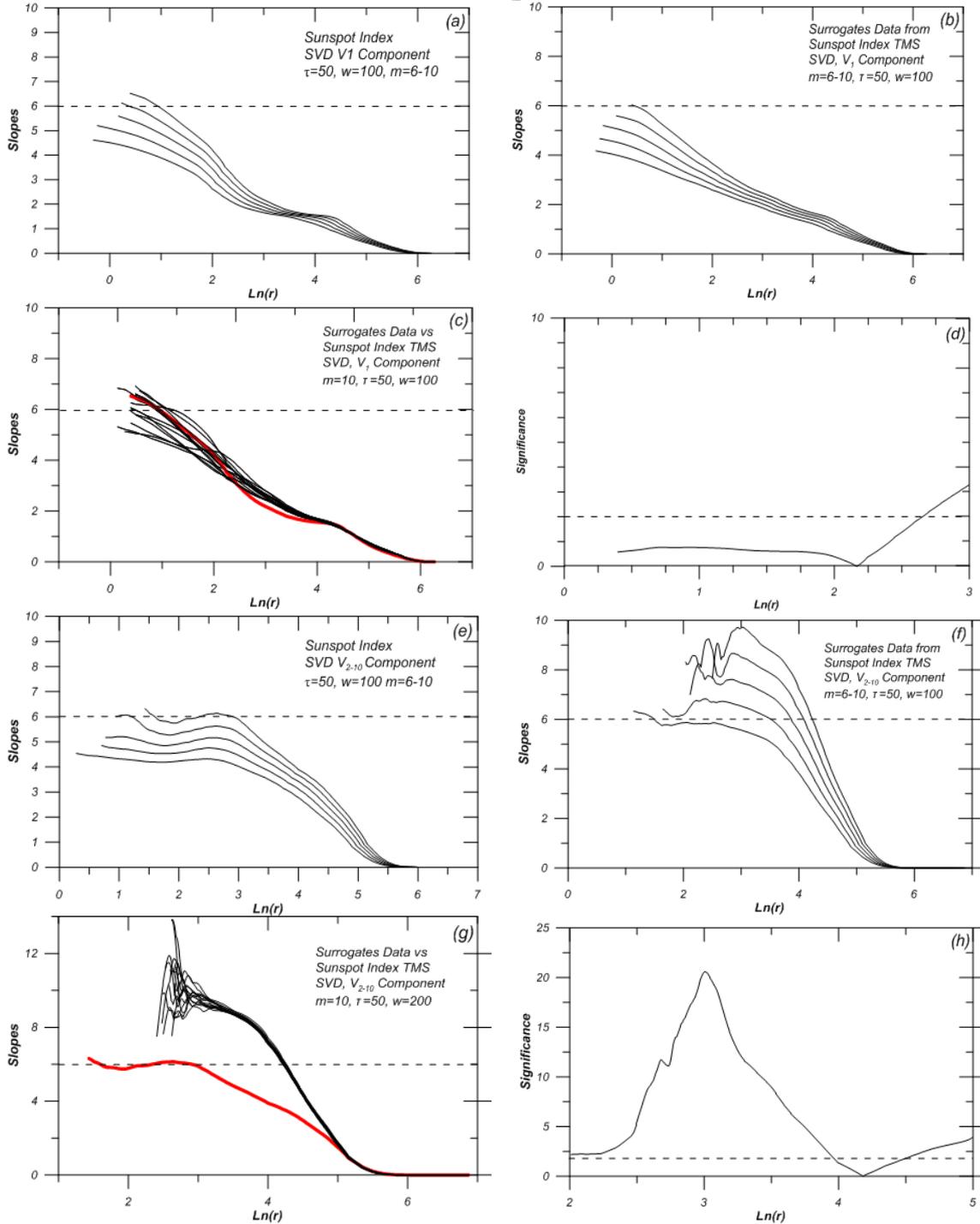

***Figure 10:*** ***(a)*** *Slopes D of the correlation integrals estimated for $V_1$ SVD component.* ***(b)*** *Slopes D of the correlation integrals estimated for the Surrogate timeseries.* ***(c)*** *Slopes of the correlation integrals of the $V_1$ SVD component and its thirty (30) surrogate estimated for delay time $\tau=50$ and for embedding dimension $m=10$, as a function of Ln(r).* ***(d)*** *The Significance of the Statistics for the $V_1$ SVD component timeseries and its 30 Surrogates.* ***(e)*** *Slopes D of the correlation integrals estimated for $V_{2-10}$ SVD component.* ***(f)*** *Slopes D of the correlation integrals estimated for the Surrogate timeseries.* ***(g)*** *Slopes of the correlation integrals of the $V_{2-10}$ SVD component and its thirty (30) surrogate, estimated for delay time $\tau=50$ and for embedding dimension $m=10$, as a function of Ln(r).* ***(h)*** *The Significance of the Statistics for the $V_1$ SVD component timeseries and its 30 Surrogates.***3.6**

# Figure 11

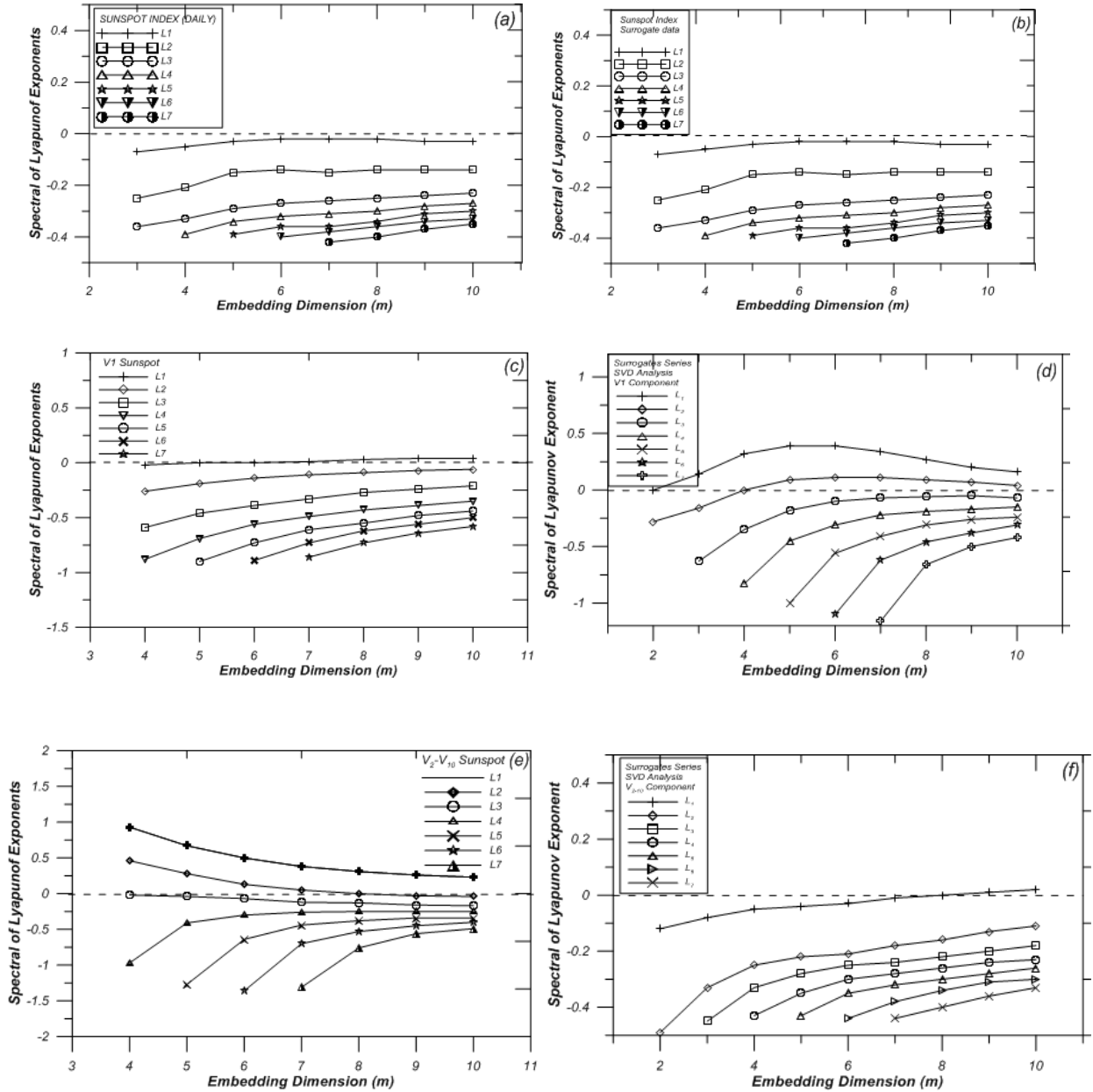

***Figure 11:*** (***a***) *The spectrum of the Lyapunov exponents $L_i$, i=1-7 as a function of embedding dimension for the Sunspot timeseries* (***b***) *The spectrum of the Lyapunov exponents $L_i$, i=1-7 as a function of embedding dimension for the Surrogate data.* (***c***) *The spectrum of the Lyapunov exponents $L_i$, i=1-7 as a function of embedding dimension for the $V_1$ SVD component* (***d***) *The spectrum of the Lyapunov exponents $L_i$, i=1-7 as a function of embedding dimension for the Surrogate data.* (***e***) *The spectrum of the Lyapunov exponents $L_i$, i=1-7 as a function of embedding dimension for the $V_{2-10}$ SVD component* (***f***) *The spectrum of the Lyapunov exponents $L_i$, i=1-7 as a function of embedding dimension for the Surrogate data.*

### 3.6 Determination of the spectra of the Lyapunov exponents

In this section we present the estimated spectra of Lyapunov exponents for the original timeseries of the sunspot index and its SVD components. Fig.11a shows the spectrum $L_i, i = 1 - 7$ of the Lyapunov exponents estimated for the original timeseries of the sunspot index. As we can observe in Fig.11a there is no positive Lyapunov exponent, while the largest one approaches the value of zero, from negative values. Fig.11b is similar to Fig.11a but for the surrogate data corresponding to the original timeseries. The similarity of the Lyapunov exponent spectra between the original signal and its surrogate data is obvious. The rejection of the null hypothesis is impossible as the significance of the discriminating statistics was estimated to be lower than two sigmas. Fig.11[c-d] is similar to the previous figures but for the $V_1$ SVD component of the original timeseries. Now, there is strong differentiation from surrogate data, as the significance of the statistics obtains values much higher than two sigmas, while the largest Lyapunov exponent obtains zero value. Finally in Fig.11[e-f], we present the Lyapunov exponent spectrum for the $V_{2-10}$ SVD component (Fig.10e) and its surrogate data (Fig.10f). In the case of the $V_{2-10}$ SVD component is also observed a strong discrimination from the surrogate data, as the significance of the statistics was found much higher than two sigmas, while the largest Lyapunov exponent was estimated to be clearly positive.

**Table 1**

| | Sunspot TMS | V1 Component | V2-10 Component |
|---|---|---|---|
| $\Delta\alpha = \alpha_{max} - \alpha_{min}$ | 1.752 | 1.113 | 1.940 |
| $\Delta(D_q)$ | 1.367 | 0.234 | 1.517 |
| q_sen | 0.368 | 0.055 | 0.407 |
| q_stat | 1.53 ± 0.04 | 1.40±0.08 | 2.12 ± 0.20 |
| q_rel (C(τ)) | 5.672±0.127 | 29.571±0.794 | 4.115±0.134 |
| q_rel (I(τ)) | 2.522±0.044 | 5.255±0.308 | 2.426±0.054 |
| L1 | ≈0 | ≈0 | >0 |
| Li, (i>2) | <0 | <0 | <0 (L2>0) |
| D (cor. Dim.) | >8 | >6 | ≈6 |

**Table 1:** *Summarize parameter values of solar dynamics: From the top to the bottom we show: changes of the ranges Δα, Δ($D_q$) of the multifractal profile. The q-triplet (q_sen, q_stat, q_rel of Tsalli's. The values of the maximum Lyapunov exponent (Li), the next Lyapunov exponent and the correlation dimension (D).*

## 4. Summary of Data Analysis

In this study we used the SVD analysis in order to discriminate the dynamical components, underlying the sunspot index timeseries. After this we applied an extended algorithm for the nonlinear analysis of the original sunspot index timeseries, its $V_1$ (first) SVD component and the signal $V_{2-10}$ composed from the sum of the higher SVD components. The analysis was expanded to the estimation of: a) Flatness coefficients as a measure of Gaussian, non-Gaussian dynamics, b) The $q$ – triplet of Tsallis non-extensive statistics, c) The correlation dimension, d) The Lyapunov exponent spectrum, e) The spectrum of the structure function scaling exponent, f) Power spectra of the signals.

The results of data analysis presented in section 3 are summarized as follows:

- Clear distinction was everywhere observed between two dynamics: a) the solar dynamics underlying the first ($V_1$) and b) the dynamics underlying the ($V_{2-10}$) SVD component of the sunspot index timeseries.

- The non Gaussian and non-extensive statistics were found to be effective for the original sunspot index as well as its SVD components $V_1, V_{2-10}$ of the sunspot, indicating non-extensive solar dynamics.

- The Tsallis $q-$triplet ($q_{sen}, q_{stat}, q_{rel}$) was found to verify according to the expected scheme $q_{sen} \leq 1 \leq q_{stat} \leq q_{rel}$. Moreover the Tsallis $q-$triplet estimation showed clear distinction between the dynamics underlying the SVD analyzed sunspot signal as follows: $q_k(V_1) < q_k(original) < q_k(V_{2-10})$, for all the $k \equiv (sen, stat, rel)$.

- The multifractal character was verified for the original signal and its SVD components. Also, the multifractality was found to be intensified as we pass from the $V_1$ to the $V_{2-10}$ SVD component in accordance to the relation: $0 < \Delta a(V_1) < \Delta a(V_{2-10})$, $0 < \Delta Dq(V_1) < \Delta Dq(V_{2-10})$, where $\Delta a = a_{max} - a_{min}$, and $\Delta Dq = D_{q=-\infty} - D_{q=+\infty}$.

- Efficient agreement between $D_q$ and $p-$model was discovered indicating intermittent (multifractal) solar turbulence which is intensified for the solar dynamics underlying the $V_{2-10}$ SVD component of the sunspot index.

- Generally the differences $\Delta a(V_i)$ and $\Delta D_q(V_i)$ increase passing from lower to higher SVD components $V_i, i = 1, 2, ..., 7$.

- The $q_{rel}$ index was estimated for two distinct relaxation magnitudes, the autocorrelation function $C(\tau)$ and the mutual information $I(\tau)$, indicating a good agreement taking into account the linear – nonlinear character of the $C(\tau)$ and $I(\tau)$ respectively.

- The correlation dimension was estimated at low values $D_{CD} = 4 - 5$ for the $V_{2-10}$ SVD component. For the $V_1$ SVD component of the original signal the correlation dimension was found to be higher than the value $\sim 10$.

- Also the null hypothesis of non-chaotic dynamics and non-linear distortion of white-noise was rejected only for the $V_{2-10}$ SVD component. For the original signal and its $V_1$ SVD component the rejection was insignificant ($s < 2$). These results indicate low dimensional deterministic solar dynamics underlying the $V_{2-10}$ SVD component and high dimensional SOC solar dynamics for the $V_1$ SVD component.

- The estimation of Lyapunov exponent spectrum showed for the $V_{2-10}$ SVD component one positive Lyapunov exponent ($\lambda_1 > 0$) discriminated from the signal of surrogate data. For the original sunspot signal and its $V_1$ SVD component the discrimination with surrogate data was inefficient. These results indicate low-dimensional and chaotic deterministic solar dynamics underlying the $V_{2-10}$ SVD component and weak chaos solar dynamics underlying to the $V_1$ SVD component related to a SOC process at the edge of chaos.

- The structure function scaling exponents spectrum $J(p)$ was estimated to values higher than the corresponding $p/3$ values of the K41 theory, for both the $V_1$ and the $V_{2-10}$ SVD components, as well as for the original signal of the sunspot index. This result indicates the intermittent (multifractal) character of the solar turbulence dissipation.
- The slopes $dJ(p)/dp$ of the scaling exponent function were found to be decreasing as the order $p$ increases. This result confirms the intermittent and multifractal character of the solar turbulence dissipation process.
- The SVD analysis of the solar dynamics in relation with the structure functions exponent spectrum $J(p)$ estimated for the SVD components showed clearly distinction of the dynamics underlying the first and the last SVD components.
- The difference $\Delta J(p)$ between the experimental and theoretical values of the scaling exponent spectrum of the structure functions was found to follow a linear profile: $\Delta J(p) = ap + b$ for the original sunspot index and its $V_1$ and the $V_{2-10}$ SVD components.
- Adding the $\Delta J(p) = ap + b$ function to the theoretically estimated $J(p)$ values by using Tsallis theory we obtain an excellent agreement of the theoretically predicted and the experimentally estimated exponent spectrum values $J(p)$.
- Noticeable agreement of Tsallis theory and the experimental estimation of the functions $f(a)$, $D(q)$, and $J(p)$ was found.

## 5. Discussion and Theoretical Interpretation of Data Analysis Results

The hydromagnetic dynamo model [85,86] can produce partly the chaotic character of the dynamics of the solar convective zone as the original MHD equations are truncated and transformed to a system of coupled nonlinear differential equations. However, the complexity of the sunspot phenomenology cannot be described faithfully by this approximation. The results of previous data analysis showed clearly:

a) The non-Gaussian and non-extensive statistical character of the solar dynamics underlying the sunspot index time series.
b) The intermittent and multifractal turbulent character of the solar plasma convection at the photosphere and the underlying convection zone.
c) The phase transition process between different dynamical profiles of the outer solar plasma (convective zone and photospheric region of the sun).
d) Novel agreement of $q-$entropy principle and the experimental estimation of solar intermittent turbulence indices: $f(a)$, $D(q)$ and $J(p)$.

These results indicate the existence of a far from equilibrium correlated with anomalous diffusion Fokker – Planck process included in the non-extensive thermodynamical and statistical theory of Tsallis [8]. The results of this study are also in agreement with experimental and theoretical results concerning the fractal geometry of the sunspot dynamics [19,20]) as well as with results obtained by Ruzmakin et al [21] concerning the anomalous diffusion and convection of the solar photosphere. The mathematical profile of solar plasma dynamics is as follows:

$$\frac{\partial^\beta P(x,t)}{\partial |t|^\beta} = -\frac{\partial}{\partial x}\big[F(x)P(x,t)\big] + D_{\beta,\gamma,q}\frac{\partial^\gamma \big[P(x,t)\big]^{2-q}}{\partial |x|^\gamma}, \qquad (5.1)$$

Such a kind of equation corresponds to the macroscopic dynamics of the solar plasma in the strange, multifractal and chaotic solar attractors, which correspond to the multiscale and intermittent solar plasma dynamics. The extended Fokker – Planck equation includes fractional space – time operators which are appropriate for mathematical modeling of anomalous diffusion processes in a fractal environment [87,88].

Fractal generalization of the know physical theory as well as fruitful results concerning self-organization, long range correlations, non-Gaussian distributions and development of non-local coherent structures in physical media in the studies of Zaslavsky [89], Tarasov [90,91,92,93,94] and in the case of the solar dynamics the fractal Fokker – Planck equation must be related to an extended modeling of dynamo process as an intermittent turbulence process of magnetized plasma. The classical mean field dynamo theory is related to the equation:

$$\frac{\partial \vec{B}}{\partial t} = \vec{\nabla} x (\vec{B}x\vec{R}x\vec{\Omega}) + a\vec{\nabla}x\vec{B} - \beta\vec{\nabla}x(\vec{\nabla}x\frac{\vec{B}}{\mu}) \qquad (5.2)$$

where $\vec{\Omega}$ is the differential rotation, $(\alpha)$ is the mean helicity, $(\beta)$ and $(\mu)$ are the turbulent diffusion and magnetic permeability respectively [12,85,86]. The intermittent turbulence with anomalous diffusion process dynamo theory must be based at the extended fractal plasma theory included anomalous magnetic transport and diffusion, magnetic percolation and magnetic Levy random walk [19,95].

For solar magnetized plasma, the fractal environment is self consistent, the plasma particles' multifractal turbulent state concerning the magnetic field anomalous diffusion, while the fractal environment causing charged particles anomalous diffusion corresponds to the multifractal topology of the magnetic field lines [21,96].

From a more extreme point of view, the fractal environment for the anomalous turbulence dissipation of magnetic field plasma particles is the fractality of the space – time itself according to Shlesinger [97], Nottale [88], Chen [52]. As we have shown in this study the intermittent turbulence of the solar convection, at the convection – photosphere zones can be explained statistically by the Tsallis entropy principle. Also the same statistical principle of non-equilibrium $q-$entropy of Tsallis statistical and thermodynamical theory can be the inner physical meaning of space time fractal Fokker – Planck equation as indicated by the results of this study in the case of solar dynamics. Now, the theoretical problem is stated as follows: How, can we relate the macroscopic phenomenology of the fractal Fokker – Planck equation to the basic plasma theory? This question carry us to the deep levels of non linear plasma theory and its physical relation with the macroscopic phenomenology of intermittent turbulence, or the phenomenology of macroscopic criticality (SOC) and chaos states [3,98,99,100,101,102,103].

The statistical description of out of equilibrium plasma states corresponds to the Bogolinbov – Born – Green – Kizkwood – Yvon (BBGKY) hierarchy of probability distributions functions in the system's phase space:

$$\vec{f} = \{f_0, f(x_i), f(x_i, x_k),...\}, \qquad (5.3)$$

where $x_i, x_j, x_k,...$ describe distinct states in the physical and (or) momentum space of plasma particles and fields $f(x_i), f(x_i, x_k),...$ are the partial probability distribution functions causing the two, three, … point dynamical correlation of the system. The BBGKY statistical hierarchy satisfies the general Liouville equation:

$$\partial_t \vec{f}(t) = \hat{L}\vec{f}(t),$$  (5.4)

where $\hat{L}$ is the Liouville operator of the system dynamics.

The BBGKY statistical hierarchy someway corresponds to the Feynman diagrams resolution of Quantum Field Theory (QFT) [104,105,105]. The application of Renormalization group (RNG) theory for the far from equilibrium plasma dynamics [56] can produce the macroscopic critical and self organized states of the system. The RNG mechanism, more than a simple mathematical description of the self similar and multiscale character of the system at its far from equilibrium critical states, includes the non local self organization physical mechanism creating long range correlations. From this point of view the $q-$entropy of Tsallis statistical theory is closely related to the far from equilibrium RNG physical mechanism of creation of long range correlations and non Gaussian statistical dynamics at the metaequilibrium stationary states of the system [8,57,60].

From this point of view the fractal Fokker – Planck equation (5.1) must be the self organizing reduction of infinite dimensional dynamics to a low dimensional chaotic dynamics or correlated SOC dynamics at the macroscopic fixed points of the metaequilibrium states. Chang [55] indicated the possibility for a macroscopic phase transition of the dynamics between distinct critical points. This is the physical meaning of the differentiation of Tsallis $q-$triplet indices and the other characteristics as the Lyapunov spectrum, the correlation dimension or the structure function exponent spectrum and intermittent turbulence characteristics, as we have shown previously in this study by the solar sunspot data analysis.

That implementation of SVD analysis in the sunspot index showed clearly the existence of two distinct solar dynamics underlying the $V_1$ and the $V_{2-10}$ SVD components of the sunspot signal. These dynamics correspond to distinct metaequilibrium stationary solar plasma states, one with SOC character and one of low dimensional chaos profile. Finally, we conclude the existence of two or more strange multifractal attractors with weak (SOC) or strong chaotic profile.